\documentclass[journal]{IEEEtran}
\usepackage{stfloats}

\usepackage{enumerate}
\usepackage{algorithm,algpseudocode}
\usepackage{setspace,amsmath,latexsym,cite,amssymb,epsfig,amsfonts}
\usepackage{url,cite}
\usepackage{graphicx}
\usepackage{psfrag}
\usepackage{footmisc}
\usepackage{multirow}
\usepackage{color}
\usepackage{multicol}
\usepackage{mathtools}
\usepackage{subcaption}
\usepackage{graphicx}
\usepackage{amssymb}
\usepackage{amsmath}
\usepackage{epstopdf}
\usepackage{geometry}
\usepackage{float}
\usepackage{balance}
\usepackage{makecell}
\usepackage{changepage}

\algnewcommand\algorithmicforeach{\textbf{for}}
\algdef{S}[FOR]{For}[1]{\algorithmicforeach\ #1\ \algorithmicdo}

\twocolumn

\makeatother

        \makeatletter
        \def\fps@eqnfloat{!t}
        \def\ftype@eqnfloat{4}
        
        \newenvironment{eqnfloat*}
               {\@dblfloat{eqnfloat}}
               {\end@dblfloat}
        \makeatother
\allowdisplaybreaks[4]

\title{\huge Joint Offloading and Resource Allocation for Hybrid Cloud and Edge Computing in SAGINs: A Decision Assisted Hybrid Action Space Deep Reinforcement Learning Approach}
\author{Chong Huang, \IEEEmembership{Member, IEEE}, Gaojie Chen, \IEEEmembership{Senior Member, IEEE}, Pei Xiao, \IEEEmembership{Senior Member, IEEE},\\Yue Xiao, \IEEEmembership{Member, IEEE}, Zhu Han, \IEEEmembership{Fellow, IEEE}, and Jonathon A. Chambers, \IEEEmembership{Fellow, IEEE}
\thanks{This work was supported in part by the U.K. Engineering and Physical Sciences Research Council under Grant EP/P03456X/1 and EP/X013162/1, partially supported by NSF CNS-2107216, CNS-2128368, CMMI-2222810, ECCS-2302469, US Department of Transportation, Toyota and Amazon.}
\thanks{C. Huang is with School of Flexible Electronics (SoFE) \& State Key Laboratory of Optoelectronic Materials and Technologies, Sun Yat-sen University, Guangdong, China, and also with 5GIC \& 6GIC, Institute for Communication Systems (ICS), Home for 5GIC \& 6GIC, University of Surrey, Guildford, GU2 7XH, United Kingdom. Email: chong.huang@surrey.ac.uk.}
\thanks{G. Chen is with School of Flexible Electronics (SoFE) \& State Key Laboratory of Optoelectronic Materials and Technologies, Sun Yat-sen University, Guangdong, China. Email: gaojie.chen@surrey.ac.uk. (Corresponding author: G. Chen)}
\thanks{P. Xiao is with 5GIC \& 6GIC, Institute for Communication Systems (ICS), Home for 5GIC \& 6GIC, University of Surrey, Guildford, GU2 7XH, United Kingdom. Email: p.xiao@surrey.ac.uk.}
\thanks{Y. Xiao is with the National Key Laboratory of Science and Technology on Communications, University of Electronic Science and Technology of China, Chengdu, 611731, China. Email: xiaoyue@uestc.edu.cn.}
\thanks{Z. Han is with the Department of Electrical and Computer Engineering at the University of Houston, Houston, TX 77004 USA, and also with the Department of Computer Science and Engineering, Kyung Hee University, Seoul, South Korea, 446-701. Email: hanzhu22@gmail.com.}
\thanks{Jonathon A. Chambers is with the School of Engineering, University of Leicester, Leicester, LE1 7RU, United Kingdom. Email: jonathon.chambers@leicester.ac.uk.}
}

\begin{document}
\captionsetup[figure]{name={Fig.},labelsep=period}

\begin{singlespace}
\maketitle
\end{singlespace}

\thispagestyle{empty}
\begin{abstract}
In recent years, the amalgamation of satellite communications and aerial platforms into space-air-ground integrated network (SAGINs) has emerged as an indispensable area of research for future communications due to the global coverage capacity of low Earth orbit (LEO) satellites and the flexible Deployment of aerial platforms. This paper presents a deep reinforcement learning (DRL)-based approach for the joint optimization of offloading and resource allocation in hybrid cloud and multi-access edge computing (MEC) scenarios within SAGINs. The proposed system considers the presence of multiple satellites, clouds and unmanned aerial vehicles (UAVs). The multiple tasks from ground users are modeled as directed acyclic graphs (DAGs). With the goal of reducing energy consumption and latency in MEC, we propose a novel multi-agent algorithm based on DRL that optimizes both the offloading strategy and the allocation of resources in the MEC infrastructure within SAGIN. A hybrid action algorithm is utilized to address the challenge of hybrid continuous and discrete action space in the proposed problems, and a decision-assisted DRL method is adopted to reduce the impact of unavailable actions in the training process of DRL. Through extensive simulations, the results demonstrate the efficacy of the proposed learning-based scheme, the proposed approach consistently outperforms benchmark schemes, highlighting its superior performance and potential for practical applications.
\end{abstract}


\section{Introduction}
\subsection{Background}
With the development of wireless communications, satellite networks have drawn significant attention in wireless communications due to their global coverage and services for ground users \cite{9508471}. Due to the various advantages of satellite communication, including scalability, high bandwidth capabilities, and the ability to reach remote and underserved areas, it plays a crucial role in various applications such as telecommunication services, broadcasting, navigation systems, and weather monitoring. In recent years, a number of ambitious satellite projects are underway, reshaping the future landscape of communication. SpaceX is one of the most notable participants in this arena \cite{McDowell_2020}, the company's Starlink project aims to establish a satellite network to provide communication services globally, especially in remote areas. Another key player is OneWeb \cite{foreman2017large}, which seeks to construct a satellite constellation to provide enhanced broadband connectivity worldwide. In addition, satellite initiatives like those from SES Networks (formerly known as O3b) and Telesat \cite{8473417} are also advancing, with their objective being to offer high-speed and low-latency internet access services with global coverage. Hence, it is evident that satellite communication has become an integral part of future communication systems. To explore the benefits of integrating satellite communication, many recent studies have investigated its implementation across various domains such as the Internet of Things (IoT) \cite{9526866}, cognitive communications \cite{8253483} and edge computing \cite{9048610}.

Moreover, considering the stringent requirements of the fifth-generation (5G) and sixth-generation (6G) in terms of communication latency and spectral efficiency, integrating terrestrial wireless networks with low Earth orbit (LEO) satellites has emerged as a highly promising solution in wireless communications \cite{9502642,9970355}. To analyze the performance in LEO satellite communications, a user grouping method was designed to enhance the communication rate in massive multiple-input-multiple-output (MIMO) systems \cite{9110855}. In \cite{9097410}, a deep learning-based algorithm was proposed to improve the drain efficiency in LEO satellite communications. To reduce the error in channel estimation of LEO satellite communications, a deep learning-based approach was designed to extract the channel features and output predictions in massive MIMO satellite communications \cite{9439942}. In \cite{9960741}, collaborative LEO satellites were introduced to realize secure and green transmissions in IoT.

Recently, the space-air-ground integrated network (SAGIN) has gained significant attention as an advanced communication framework that seamlessly integrates satellites, high-altitude platforms (HAPs), unmanned aerial vehicles (UAVs) and ground-based networks, providing improved connectivity and enhanced communication capabilities \cite{8368236}. Satellites play a vital role in SAGINs by providing ubiquitous global coverage, ensuring communication accessibility even in remote and underserved regions. HAPs and UAVs offer unparalleled flexibility, mobility, and rapid deployment capabilities. Ground-based networks serve as a robust backbone, providing localized connectivity and essential infrastructure support. Through the integration of these components, SAGIN enables efficient and reliable communication across diverse domains, including target localization \cite{9380358}, the Internet of Vehicles (IoV) \cite{9369409}, cognitive communications \cite{10287142} and IoT networks \cite{9626560}. The authors in \cite{9380358} investigated ocean target localization in SAGIN with multi-UAV cooperative assistance. To enhance the average data to overhead ratio, UAVs were utilized in SAGIN to improve the transmission quality for the IoV \cite{9369409}. A learning-based UAV trajectory and reconfigurable intelligent surface (RIS) phase shift design was proposed to enhance the secrecy performance in UAV-RIS enabled cognitive SAGIN networks \cite{10287142}. In \cite{9626560}, UAVs were adopted to enhance the wireless power transmission in the SAGIN-IoT. This integration of space-air-ground resources unlocks unprecedented potential for advanced communication systems in a wide range of environments.

On the other hand, multi-access edge computing (MEC) is a revolutionary concept in mobile communications that aims to meet the increasing demand for low-latency and high-bandwidth applications and services \cite{7901477}. It achieves this by shifting computational tasks from the cloud to the network edge, thereby bringing computing resources closer to mobile devices. Various MEC frameworks have been studied in current works \cite{8488502,9133107,9893789}. In \cite{8488502}, a convex optimization-based method was proposed to reduce the energy consumption in MEC frameworks. To reduce the latency in MEC, the authors of \cite{9133107} applied non-convex optimization algorithms to MEC networks. In \cite{9893789}, an alternative optimization (AO)-based algorithm was proposed to minimize the latency of MEC in millimeter-wave communications.

\subsection{Related Work}\label{sec:RW}
HAPs and UAVs can offer dynamically allocated computing resources to meet the evolving needs of users and computing tasks in MEC \cite{Liu22,9725258,10075438,10106141}. Therefore, many current works have exploited the benefits of the integration of UAV-assisted communications and MEC. In \cite{Liu22}, deep reinforcement learning (DRL) algorithms were utilized to optimize the virtual machine configuration and UAV's trajectory to reduce the latency in MEC. The resource allocation and UAV's trajectory were adjusted to reduce the energy consumption in MEC via a multi-agent DRL algorithm \cite{9725258}. The authors of \cite{10075438} proposed a DRL-based optimization of UAV's trajectory, directed acyclic graph (DAG) scheduling and service function deployment to enhance the success rate in path finding. To improve the users’ satisfaction, the genetic algorithm and the K-means algorithm were adopted to optimize the UAV's trajectory and task offloading in MEC \cite{10106141}. Although these studies recognize the importance of UAVs in MEC, they do not consider cloud servers and satellite communications which could provide global coverage.

Furthermore, due to the ability of satellites to provide global coverage and access services in remote areas, many recent studies have investigated the advantages of incorporating satellite communication within the MEC framework \cite{9184934,9383778,9344666,9978924,9894082}. In \cite{9184934}, the Lagrange multiplier method and DRL algorithm were utilized to optimize the resource allocation in satellite-aided vehicle-to-vehicle networks to reduce the overall latency in MEC. To reduce the computational cost in MEC, the authors of \cite{9383778} used sequential fractional programming and the Lagrangian dual decomposition method to optimize the computation offloading and resource allocation in the satellite-terrestrial communications. In \cite{9344666}, the authors utilized an alternating direction method of multipliers to minimize the energy cost in a hybrid cloud and MEC framework. To obtain the tradeoff between the latency and computational cost in MEC, convex optimization algorithms were adopted to optimize the resource allocation in SAGIN \cite{9978924}. In \cite{9894082}, a greedy strategy was proposed to optimize the resource allocation in MEC to reduce the energy consumption and latency in LEO satellite networks. However, the dynamic access capability of HAPs and UAVs in SAGIN, cloud server selection and multi-task offloading were not considered in the above works.

In addition, by combining the strengths of satellites and UAVs, SAGIN offers a unique advantage in MEC \cite{8436043}. Consequently, to capitalize on the advantages of SAGINs, some recent researches have studied the performance of MEC under the SAGIN architecture \cite{9515574,9928786,10024305}. In \cite{9515574}, the joint offloading and resource allocation optimization was adopted to reduce the energy consumption in a satellite-HAP-ground framework. The authors of \cite{9928786} utilized DRL and convex optimization algorithms to optimize the computation resources in MEC to enhance the energy backup and average communication rate in SAGIN. In \cite{10024305}, the energy consumption was reduced by using a DRL algorithm to jointly optimize the offloading and resource allocation for MEC in SAGIN. However, these works fail to consider the dynamic grouping and access capabilities of UAVs within the aerial layer, as well as the substantial computational power provided by cloud servers. To reduce the energy cost for MEC in IoT networks, the joint UAV placement, MEC resource allocation and offloading were optimized via an approximation algorithm in \cite{9420280}. In \cite{9709222}, approximation algorithms were utilized to adjust the UAV placement and resource allocation for UAV-assisted communications to enhance the quality of experience (QoE) for the virtual reality applications in MEC. Although these works considered the dynamic access capabilities of UAVs in MEC, the dynamic grouping, LEO satellites and clouds are not included in their contributions.

\subsection{Motivation and Contributions}
In this study, we explore the integration of hybrid multi-cloud service and MEC in SAGINs. To the best of the authors' knowledge, we address a research gap as the current literature does not address the dynamic access capability of UAVs, multi-satellite access in hybrid cloud environments, cloud service selection and MEC resource allocation simultaneously. The contributions of our work and a comparative analysis with existing literature are summarized in Table \ref{tab:table1}. Our investigation also encompasses multi-task scheduling based on DAG and dynamic pairing of UAV-ground users. Furthermore, we extend our scope to cover partial offloading, task dependency and cloud selection in MEC. To effectively reduce energy consumption and latency in MEC, a joint optimization of MEC resources, UAV trajectories, and cloud server selection becomes crucial. In addition, traditional deep reinforcement learning struggles to explore vast action spaces within such complex environments without any form of human assistance. To tackle this challenge, we propose a decision-assisted hybrid multi-agent soft actor-critic (SAC) algorithm capable of addressing this complicated optimization problem. The main contributions of this work are summarized as follows.

\begin{table*}[t]
 \caption{COMPARISON BETWEEN OUR WORK AND EXISTING WORKS}
  \centering
  \scalebox{0.9}{
  \begin{tabular}{|c|c|c|c|c|c|c|c|c|c|c|c|c|c|}
  \hline
  Novelty& Our Work&\makecell{\cite{9928786}\\2023}&\makecell{\cite{9978924}\\2023}&\makecell{\cite{10024305}\\2023}&\makecell{\cite{10106141}\\2023}&\makecell{\cite{10075438}\\2023}&\makecell{\cite{9894082}\\2022}&\makecell{\cite{9515574}\\2022}&\makecell{\cite{9725258}\\2022}&\makecell{\cite{Liu22}\\2022}&\makecell{\cite{9344666}\\2021}&\makecell{\cite{9383778}\\2021}&\makecell{\cite{9184934}\\2021}\\
  \hline
  Satellite&\checkmark&\checkmark&\checkmark&\checkmark& & &\checkmark&\checkmark& & &\checkmark&\checkmark&\checkmark\\
  \hline
  Multiple satellites&\checkmark&\checkmark&\checkmark&\checkmark& & &\checkmark& & & &\checkmark&\checkmark&\\
  \hline
  Satellite coverage&\checkmark&&&\checkmark& & &\checkmark& & & &\checkmark& &\\
  \hline
  UAV/HAP&\checkmark&\checkmark& &\checkmark&\checkmark&\checkmark& &\checkmark&\checkmark&\checkmark& & &\\
  \hline
  UAV trajectory&\checkmark& & & & &\checkmark& & &\checkmark&\checkmark& & &\\
  \hline
  Cloud service&\checkmark& &\checkmark&& & & & & & &\checkmark& &\\
  \hline
  Cloud service selection&\checkmark& & & & & & & & & & & &\\
  \hline
  User pairing&\checkmark& &  & &\checkmark&& &\checkmark& & & & &\\
  \hline
  Partial offloading&\checkmark&\checkmark&\checkmark& & & & &\checkmark&\checkmark& & &\checkmark&\\
  \hline
  Multiple tasks&\checkmark& & &\checkmark& &\checkmark& & & & & & &\\
  \hline
  Task dependency&\checkmark& & &\checkmark& &\checkmark& & & & & & &\\
  \hline
  Cost minimization&\checkmark&\checkmark&\checkmark&\checkmark& & &\checkmark&\checkmark&\checkmark& &\checkmark&\checkmark&\\
  \hline
  Latency minimization&\checkmark&&\checkmark& & & &\checkmark& & &\checkmark & & &\checkmark\\
  \hline
  DRL algorithm&\checkmark&&&\checkmark&&\checkmark&&&\checkmark&\checkmark& & &\\
  \hline
  Hybrid action space&\checkmark&&& & & && & & & & &\\
  \hline
  Decision assistant&\checkmark&&& & & && & & & & &\\
  \hline
 \end{tabular}
 }
 \label{tab:table1}
 \end{table*}

\begin{itemize}
  \item We propose the integration of hybrid multi-cloud and MEC in SAGINs. Our framework considers the dynamic access capability of multiple UAVs in MEC to meet the needs of ground users. Besides, we consider the existence of multiple satellites, inter-satellite communications, cloud server selection and also account for varying coverage times for each LEO satellite in SAGINs.

  \item In the framework of MEC, we model multiple tasks using DAG and consider challenges such as user pairing, partial offloading, task dependency, satellite coverage, and cloud selection. We formulate two optimization problems: 1.) minimizing energy consumption with a latency constraint; 2.) minimizing average latency with an energy consumption constraint.

  \item To tackle the joint optimization problem of offloading and resource allocation for MEC, we utilize the multi-agent DRL algorithm in this work. This DRL-based algorithm enables agents to explore the environment and learn diverse and robust policies. To handle the hybrid discrete and continuous action space problem, we introduce a hybrid reinforcement learning framework which separates discrete and continuous actions among different agents to facilitate collaborative training. This framework utilizes use off-policy data and optimizes discrete and continuous actions simultaneously for the hybrid policy. Moreover, a decision-assisted DRL framework is adopted to reduce the negative impact of unavailable actions in the training process for deep neural networks to enhance the convergence performance of the proposed hybrid action space DRL. Therefore, we have developed a pre-trained algorithm capable of adapting to complex dynamic environments within the proposed heterogeneous space-air-ground network, with rapid decision-making ability.

  \item Simulation results demonstrate that the proposed DRL-based algorithm effectively leverages multi-agent collaboration to approximate optimal solutions. By building a global model based on rewards, the agents generate a joint optimization strategy, outperforming benchmark methods and achieving superior performance in the considered tasks.
\end{itemize}

The remainder of this paper is structured as follows: Section \ref{sec:sm} introduces the system model and the optimization problems. Section \ref{sec:SAC} presents the proposed hybrid DRL algorithm for optimizations in two cases. The simulation results are presented in Section \ref{sec:sim}. Finally, Section \ref{sec:con} concludes the work.

\section{System Model and Problem Formulation} \label{sec:sm}
\begin{figure}[t!]
  \centering
  \centerline{\includegraphics[width=0.75\textwidth]{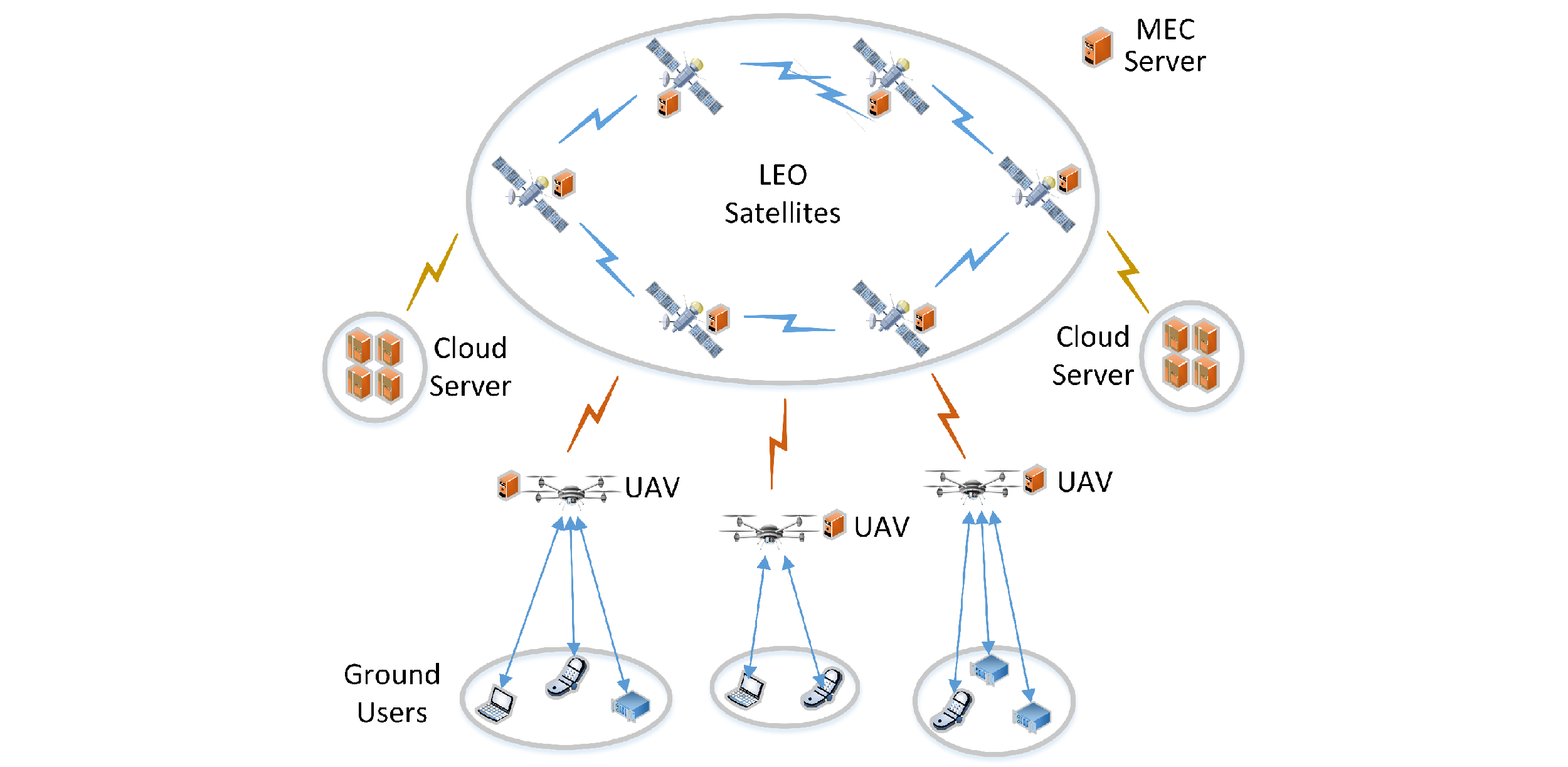}}
 \caption{System model of a hybrid cloud and MEC SAGIN.} \label{fig:SM}
\end{figure}
As shown in Fig. \ref{fig:SM}, we consider the adoption of heterogeneous MEC offloading in SAGIN to provide global communication and computing services to ground users in remote areas. The SAGIN consists of $M$ ground users $G_m$, where $m \in \mathcal M = \{1, 2, ..., M\}$. In the aerial layer, there are $N$ UAVs $U_n$, where $n \in \mathcal N = \{1, 2, ..., N\}$. In the satellite layer, we assume $L$ LEO satellites $S_l$, where $l \in \mathcal L = \{1, 2, ..., L\}$. LEO satellites can access to $K$ cloud servers $CS_k$, where $k \in \mathcal K = \{1, 2, ..., K\}$. We assume that each LEO satellite and UAV is equipped with a server, enabling them to provide MEC services to ground users. Furthermore, considering the high-speed movement of LEO satellites, the communication time between UAVs and LEO satellites is limited by the coverage time of the LEO satellites. Moreover, since UAVs require dynamic trajectory planning to optimize the communication rate with ground users, we assume the UAV $U_n$'s 3D coordinate\footnote{In this work, the UAV flight is constrained by its flying speed and flying zone. Given the limited satellite access time, the energy limitations of UAVs are not within the scope of this study.} is $q_{n}(t) = \{x_{n}(t), y_{n}(t), z_{n}(t)\}$ at time slot $t$.

\subsection{Communication Model}
In this work, we assume that each ground user can only access to one UAV to upload its tasks for edge computing at a give time slot. The channels between UAVs and ground users follow Rician fading, the channel coefficient $h_{m n}$ between the ground user $G_m$ and UAV $U_n$ can be expressed as
\begin{equation}\small\label{eq:hmn}
h_{m n}= \sqrt{\frac{\xi}{\xi+1}} \bar{H}_{m n}+\sqrt{\frac{1}{\xi+1}} \hat{H}_{m n},
\end{equation}
where $\xi$ denotes the Rician factor, $\hat{H}_{m n}=\hat{g}_{m n} d_{m n}^{-{\alpha_N} / 2}$ and $\bar{H}_{m n}=\bar{g}_{m n} d_{m n}^{-{\alpha_L} / 2}$ are the non-line-of-sight (NLoS) component and the line-of-sight (LoS) component in Rician fading, respectively. $d_{m n}$ is the distance between ground user $G_m$ and UAV $U_n$, $\alpha_L$ and $\alpha_N$ are the path loss exponents for LoS and NLoS in Rician fading, respectively. $\hat{g}_{m m}$ is the NLoS channel coefficient represented by a zero-mean unit-variance Gaussian fading channel, $|\bar{g}_{m n}| = 1$ is the LoS channel coefficient. Thus, the uplink transmission rate for the offloading between $G_m$ and $U_n$ is given by
\begin{equation}\small\label{eq:Ratemn}
R_{m n}= B_{m n} {\rm{log_{2}}} \left(1 + \frac{P_G |h_{m n}|^2} { \sum_{a=1,a \neq m}^{M}P_G |h_{a n}|^2 + {{\sigma}_{n}^2} }   \right),
\end{equation}
where $B_{m n}$ denotes the channel bandwidth between $G_m$ and $U_n$, $P_G$ is the transmit power for all $G_m$, $\sum_{a=1,a \neq m}^{M}P_G |h_{a n}|^2$ denotes the interference from other ground users, ${{\sigma}_{n}^2}$ is the variance of the additive-white-Gaussian-noise (AWGN) at $U_n$.

Moreover, we assume that one UAV can only access to one LEO satellite at a given time slot. The channel coefficient between a UAV $U_n$ and a LEO satellite $S_l$ is given by \cite{9726800}
\begin{equation}\small\label{eq:G_s}
\bar{h}_{n l}=\frac{\sqrt{\delta_l}\lambda}{4 \pi d_{n l}} e^{j {\epsilon}},
\end{equation}
where $\delta_l$ denotes the beam gain, $\lambda$ denotes the wavelength, $d_{n l}$ is the distance between $U_n$ and $S_l$, $\epsilon$ is the antenna phase. Moreover, we consider the outdated channel state information (CSI) for the channel coefficients ${h}_{n l}$ due to the transmission delay in satellite communications \cite{8822992,9535285}. Therefore, we can obtain
\begin{equation}\small\label{eq:outdatedCSI}
h_{n l}= \varepsilon \bar{h}_{n l} + \sqrt {1-\varepsilon^2} g_{n l},
\end{equation}
where $\varepsilon = \bar{J}_0 (2 \pi {\hat{f}}_{n l} T_{n l})$, $\bar{J}_0$ is the zeroth order Bessel function of the first kind, ${\hat{f}}_{n l}$ and $T_{n l}$ are the maximum Doppler frequency and the delay of the transmissions between $U_n$ and $S_l$, respectively. $g_{nl}$ is a complex Gaussian random variable which having the same variance as $\bar{h}_{n l}$. Consequently, the uplink transmission rate for the offloading between $U_n$ and $S_l$ is given by
\begin{equation}\small\label{eq:Ratenl}
R_{n l}= B_{n l} {\rm{log_{2}}} \left(1 +  \frac{P_U |h_{n l}|^2} { \sum_{a=1,a \neq n}^{N} P_U |h_{a l}|^2 + {{\sigma}_{l}^2} }   \right),
\end{equation}
where $B_{n l}$ denotes the channel bandwidth between $U_n$ and $S_l$, $P_U$ is the transmit power for all $U_n$, $\sum_{a=1,a \neq n}^{N} P_U |h_{a l}|^2$ is the interference from other UAVs, ${\sigma}_{l}^2$ is AWGN at $S_l$.

In addition, each LEO satellite can share information from ground users with other LEO satellites via inter-satellite link (ISL). The channel rate for an ISL between LEO satellites $S_{i}$ and $S_{j}$ is given by \cite{9327501}
\begin{equation}\small\label{eq:rate_ISL}
R_{i j} = B_{i j} {\rm{log_{2}}} \left( 1 +  \frac{P_S |G_{\rm max}|^2}{\kappa \zeta B_{i j} (\frac{{4 \pi d_{i j} f_S}}{c})^2 } \right),
\end{equation}
where $B_{i j}$ denotes the channel bandwidth for ISL, $P_S$ is the transmit power for all $S_l$. $G_{\rm max} = {\rm max}_{f(i, j)} {G^d_{i j}}$ denotes the peak gain of antennas of satellite in the direction of their main lobe, where $f(i, j)$ is relative direction between LEO satellites $i$ and $j$, $d_{i j}$ denotes the distance between satellite $i$ and satellite $j$. $G^d_{i j}$ denotes the normalized gain for the ISL between $S_i$ and $S_j$. $\kappa$ is the Boltzmann constant, $\zeta$ denotes the thermal noise, $f_S$ is the carrier frequency and $c$ is the speed of light. It is noticed that the interference can be ignored in the intra-plane ISLs \cite{9327501}.

On the other hand, the LEO satellites can access to the cloud servers to offload tasks. The cloud servers are deployed on the ground and each LEO can only access to one cloud server. The transmission rate between a LEO satellite $S_l$ and a cloud server $CS_{k}$ is given by
\begin{equation}\small\label{eq:RateLEO_Cloud}
R_{l k}= B_{l k}  {\rm{log_{2}}} \left(1 +  \frac{P_S |h_{l k}|^2} { {\sigma}_{l}^2} \right),
\end{equation}
where $B_{l k}$ denotes the channel bandwidth for the transmission between $S_l$ and $CS_{k}$, $h_{l k}$ denotes the channel coefficients between satellite $S_l$ and cloud server $CS_{k}$.

\subsection{LEO Coverage Model}
\begin{figure}[t!]
  \centering
  \centerline{\includegraphics[width=0.31\textwidth]{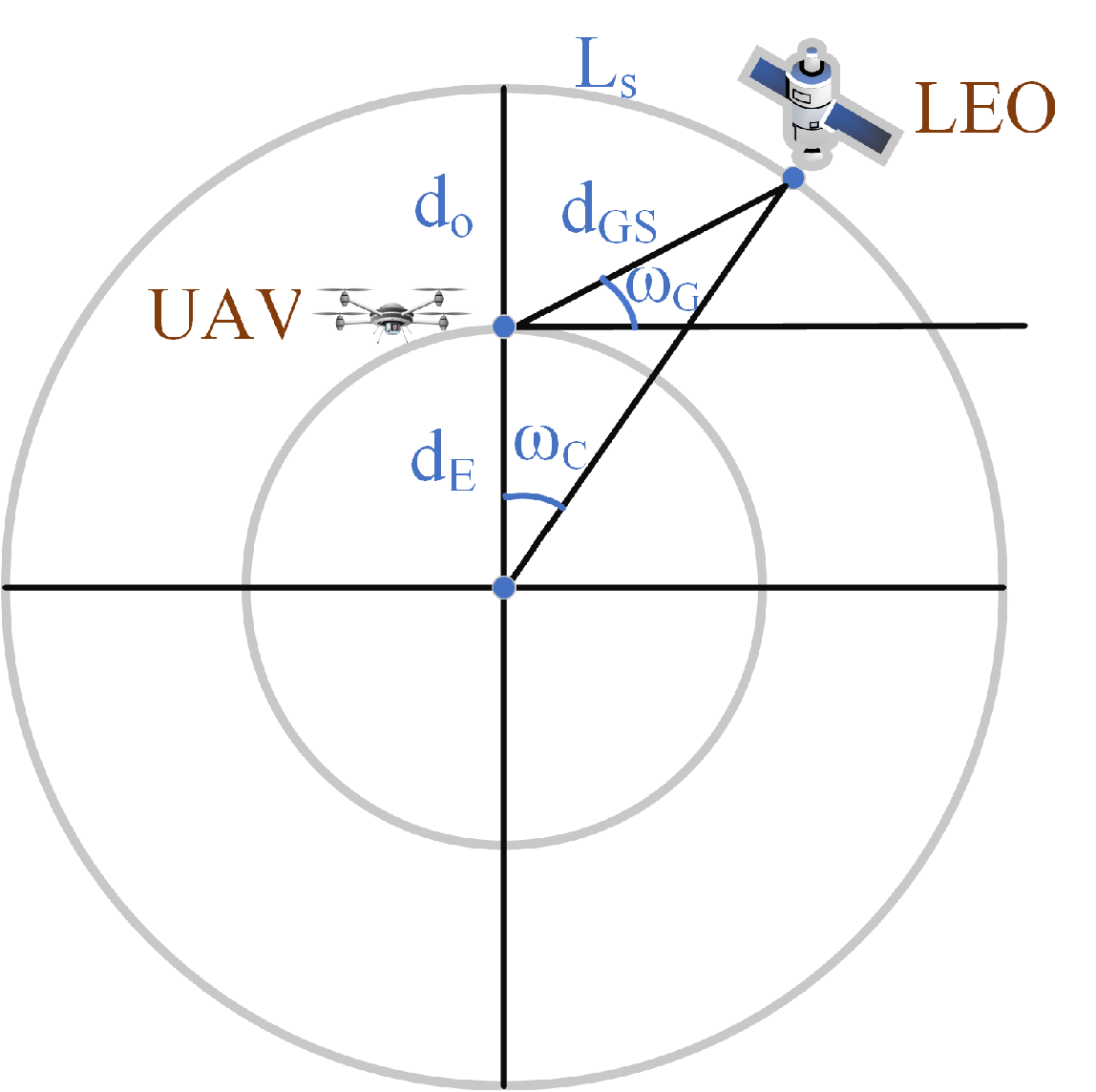}}
 \caption{The coverage provided by LEO satellites for UAVs.} \label{fig:coverage}
\end{figure}
Due to the continuous high-speed movement of LEO satellites, the window of time during which LEO satellites can provide MEC service to UAVs is limited \cite{9344666}. As shown in Fig. \ref{fig:coverage}, the angle between the UAVs and the LEO satellite can be expressed as
\begin{equation}\small\label{eq:elevationAngle}
\omega_G = \arccos \bigg(\frac{d_E + d_O}{d_{GS}} \sin \omega_C \bigg),
\end{equation}
where $d_E$ is the radius of the earth, $d_O$ is the height of the LEO satellite orbit, $d_{GS}$ denotes the distance between the UAVs and the LEO satellite, $\omega_C$ presents the angle for the coverage of the LEO satellite. Notice that $\omega_G$ is a predefined minimum elevation angle for the satellite coverage area, when the elevation angle between the satellite and UAV falls below this predetermined threshold, the satellite can no longer provide service to the relevant UAV. Thus, $\omega_C$ can be expressed as
\begin{equation}\small\label{eq:Angle2}
\omega_C = \arccos \bigg(\frac{d_E}{d_E+d_O} \cos \omega_G \bigg) - \omega_G.
\end{equation}
Then we can obtain that the arc length of the LEO satellite during the coverage time as
\begin{equation}\small\label{eq:arcLength}
L_S = 2 (d_E + d_O) \omega_C.
\end{equation}
Thus, the communication time $T_{U_n S_l}$ between $U_n$ and $S_l$ is
\begin{equation}\small\label{eq:comTime}
T_{U_n S_l} = \frac{L_S}{v_S},
\end{equation}
where $v_S$ is the orbital velocity of LEO satellites. Considering the significantly smaller values of UAV height and distance between UAVs compared to $d_E$ and $d_O$, it can be safely neglected when calculating the LEO satellite coverage time. Furthermore, considering the distance between LEO satellites, the remaining service time for each satellite varies in this work. Moreover, the ground cloud servers are located at different positions, which may be very far away from each other. Therefore, each LEO satellite provides different service time for different cloud servers. Considering that ISL is utilized in the communications between LEO satellites, if a cloud server finishes computing and there is a LEO satellite that can still access it within its service time of the UAVs, it is assumed that the cloud server can send the computation results back to the ground users.

\subsection{Computation Model}
\begin{figure}[t!]
  \centering
  \centerline{\includegraphics[width=0.37\textwidth]{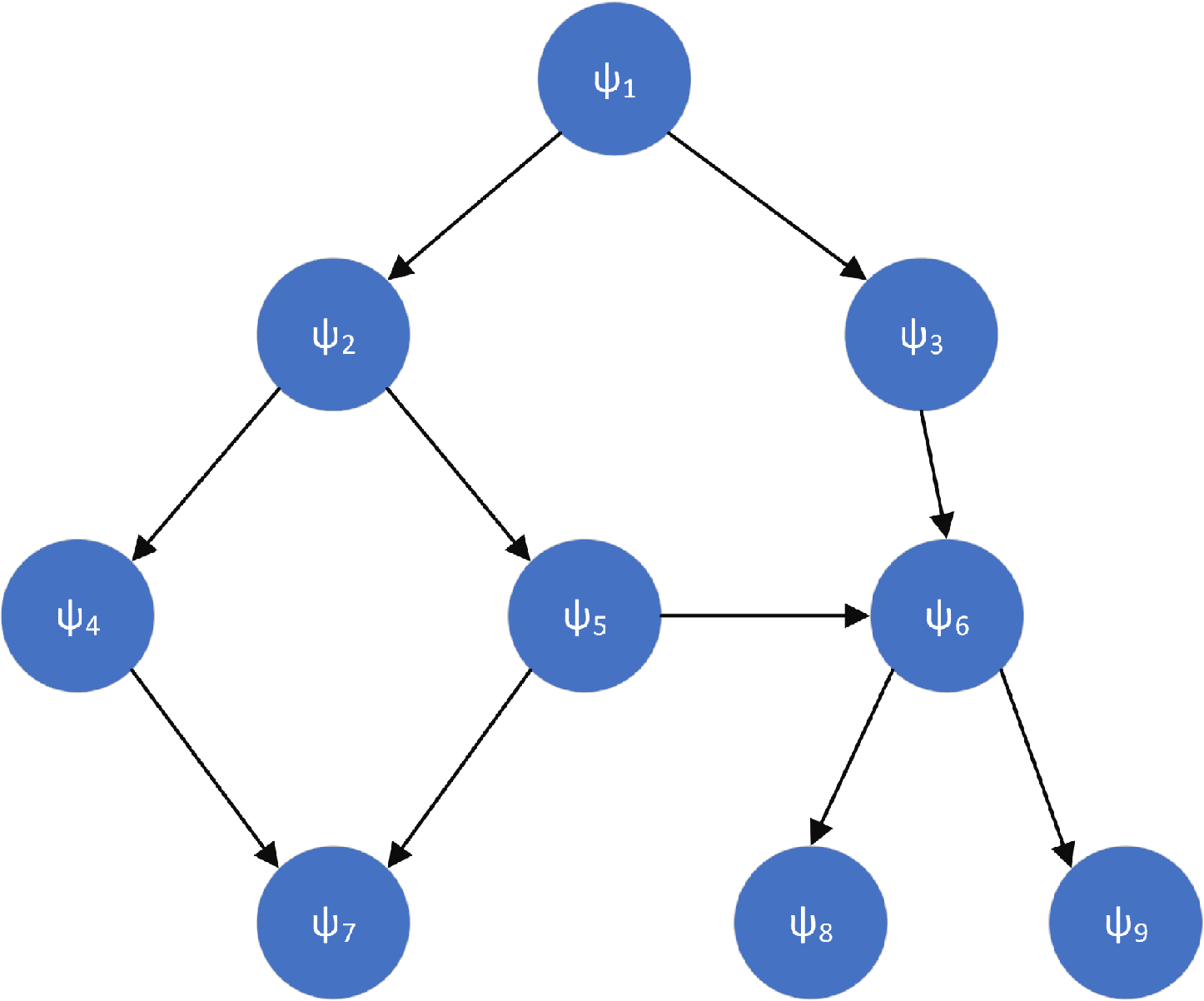}}
 \caption{DAG task dependency.} \label{fig:DAG}
\end{figure}

This work considers a scenario where all ground users are responsible for handling multiple computing tasks. These tasks can exhibit dependencies, meaning that the completion of certain tasks is required before others can proceed, such dependencies often introduce an order or sequence in which tasks must be executed. Therefore, understanding and managing these task dependencies are essential for optimizing system performance and ensuring efficient task execution. In this paper, we model the multiple tasks from a ground user as a DAG \cite{10024305,10075438}, which provide a structured representation of the dependencies and relationships among tasks, enabling efficient scheduling and resource allocation. In a DAG, each node represents a specific task, and the edges depict the dependencies between tasks. By utilizing DAGs, IoT systems can optimize task execution, ensure proper sequencing, and allocate resources effectively, leading to improved system performance and overall efficiency. For example, as shown in Fig. \ref{fig:DAG}, $\psi_1$ is the parent task of $\psi_2$ and $\psi_3$, while $\psi_2$ and $\psi_3$ represent the child tasks of $\psi_1$, respectively. The execution of a child task is contingent upon the completion of its parent tasks.

Furthermore, the MEC architecture in this paper is divided into four layers: ground users, UAVs, LEO satellites, and cloud servers. Each UAV can serve multiple ground users, but each ground user can only access to one UAV. Each satellite can serve multiple UAVs, but can only access to one cloud server. We consider partial offloading protocols in this work\footnote{In this work, each task can be partially offloaded to other MEC units, while still adhering to the task dependency in the DAG. Tasks offloaded to other MEC units need to wait for the completion of their parent tasks before they can start computation.}, where each task is bit-wise independent and can be arbitrarily allocated to other MEC units \cite{9515574}.

\subsubsection{Ground User Computing}
For a given task $\psi^m_i$ from ground user $G_m$, we assume $\psi^m_i \triangleq \{W_i, F_i\}$ where $W_i$ denotes the size of the input data, $F_i$ denotes the required central processing unit (CPU) cycles. Ground user $G_m$ has the flexibility to perform computations either locally or upload certain parts of the task to a UAV. The computation execution time of task $\psi^m_i$ at $G_m$ is
\begin{equation}\small\label{eq:comTimeGm}
T^{C}_{m, i} = \frac{\mu^i_m F_i}{f_m},
\end{equation}
where $\mu^i_m \in [0, 1]$ denotes the partial offloading factor for $G_m$, $f_m$ denotes the computation resource of $G_m$. The energy consumption of local computing at $G_m$ is expressed as
\begin{equation}\small\label{eq:energyComGm}
E^{C}_{m, i} = \iota \mu^i_m {F_i} f^2_m,
\end{equation}
where $\iota$ is the energy factor depending on the $G_m$'s architecture.

\subsubsection{UAV Computing}
When $\mu^i_m < 1$, $G_m$ offload parts of computing task $\psi^m_i$ to UAV $U_n$, we assume the pairing binary variable $z_{m, n} = 1$ \footnote{A binary variable can only be 0 or 1. In this work, we assume that all binary variables are 0 unless we assign them the value of 1 in the pairing.}, the transmission time is given as
\begin{equation}\small\label{eq:transTimeGmUn}
T^{T}_{m, n} = \frac{b (1-\mu^i_m)W_i}{R_{m n}},
\end{equation}
where $b > 1$ is the transmission overhead factor \cite{9515574}. The energy consumption of the transmission between $G_m$ and $U_n$ is expressed as
\begin{equation}\small\label{eq:energyTransGmUn}
E^{T}_{m, n, i} = P_G T^{T}_{m, n}.
\end{equation}
Furthermore, the computation execution time of task $\psi^m_i$ at $U_n$ is
\begin{equation}\small\label{eq:comTimeUn}
T^{C}_{n, i} = \frac{ \mu^i_n (1-\mu^i_m) F_i}{f_{n, i}},
\end{equation}
where $\mu^i_n \in [0, 1]$ denotes the partial offloading factor for $U_n$, $f_{n, i}$ denotes computation resource allocated to task $\psi^m_i$ from $U_n$. The energy consumption of computing at $U_n$ is expressed as
\begin{equation}\small\label{eq:energyComUn}
E^{C}_{n, i} = \iota \mu^i_n (1-\mu^i_m) {F_i} f^2_{n, i}.
\end{equation}

\subsubsection{LEO Satellite Computing}
When $\mu^i_n < 1$, $U_n$ also offloads parts of computing task $\psi^m_i$ to LEO satellite $S_l$, we assume the pairing binary variable $z_{n, l} = 1$. Nevertheless, the considerable distance between UAVs and LEO satellites introduces propagation delays that impact the communication between them. Consequently, the transmission time of computation task $\psi^m_i$ on LEO satellite $S_l$ includes the propagation delay, which can be expressed as
\begin{equation}\small\label{eq:transTimeUnSl}
T^{T}_{n, l} = \frac{d_{n l}}{c} + \frac{b (1-\mu^i_n) (1-\mu^i_m) W_i}{R_{n l}},
\end{equation}
where $\mu^i_n \in [0, 1]$ denotes the partial offloading factor for $U_n$. The energy consumption of the transmission between $U_n$ and $S_l$ is expressed as
\begin{equation}\small\label{eq:energyTransUnSl}
E^{T}_{n, l, i} = P_U \frac{b (1-\mu^i_n) (1-\mu^i_m) W_i}{R_{n l}}.
\end{equation}
Furthermore, the computation execution time of task $\psi^m_i$ at $S_l$ is
\begin{equation}\small\label{eq:comTimeSl}
T^{C}_{l, i} = \frac{ \mu^i_l (1-\mu^i_n) (1-\mu^i_m) F_i}{f_{l, i}},
\end{equation}
where $\mu^i_l \in [0, 1]$ denotes the partial offloading factor for $S_l$, $f_{l, i}$ denotes $S_l$ computation resource allocated to task $\psi^m_i$. The energy consumption of computing at $S_l$ is expressed as
\begin{equation}\small\label{eq:energyComSl}
E^{C}_{l, i} = \iota \mu^i_l (1-\mu^i_n) (1-\mu^i_m) {F_i} f^2_{l, i}.
\end{equation}
When $\mu^i_l < 1$, $S_l$ can offloads the rest of computing task $\psi^m_i$ to other LEO satellites and a cloud server. We assume $S_l$ can only offload a given task once. Therefore, if $S_l$ offloads the rest of computing task $\psi^m_i$ to another LEO satellite $s_j$, we assume the pairing binary variable $z_{l, j} = 1$, the transmission time is
\begin{equation}\small\label{eq:comTimeSj}
T^{T}_{j, i} = \frac{d_{l j}}{c} + \frac{b (1-\mu^i_l) (1-\mu^i_n) (1-\mu^i_m) W_i}{R_{l j}},
\end{equation}
where $d_{l j}$ is the distance between $S_l$ and $S_j$. The energy consumption of the transmission between $S_l$ and $S_j$ is expressed as
\begin{equation}\small\label{eq:energyTransSlSj}
E^{T}_{l, j, i} = P_S \frac{b (1-\mu^i_l) (1-\mu^i_n) (1-\mu^i_m) W_i}{R_{l j}}.
\end{equation}
Furthermore, the computation execution time of task $\psi^m_i$ at $S_j$ is
\begin{equation}\small\label{eq:comTimeSj}
T^{C}_{j, i} = \frac{ (1-\mu^i_l) (1-\mu^i_n) (1-\mu^i_m) F_i}{f_{j, i}},
\end{equation}
where $f_{j, i}$ denotes $S_j$ computation resource allocated to task $\psi^m_i$. The energy consumption of computing at $S_j$ is expressed as
\begin{equation}\small\label{eq:energyComSj}
E^{C}_{j, i} = \iota (1-\mu^i_l) (1-\mu^i_n) (1-\mu^i_m) {F_i} f^2_{j, i}.
\end{equation}

\subsubsection{Cloud Server Computing}
When $\mu^i_l < 1$ and $S_l$ offloads the rest of computing task $\psi^m_i$ to a cloud server $CS_k$, we assume the pairing binary variable $z_{l, k} = 1$, the transmission time is
\begin{equation}\small\label{eq:comTimeCSk}
T^{T}_{k, i} = \frac{d_{l k}}{c} + \frac{b (1-\mu^i_l) (1-\mu^i_n) (1-\mu^i_m) W_i}{R_{l k}}.
\end{equation}
The energy consumption of the transmission between $S_l$ and $CS_k$ is expressed as
\begin{equation}\small\label{eq:energyTransSlSj}
E^{T}_{l, k, i} = P_S \frac{b (1-\mu^i_l) (1-\mu^i_n) (1-\mu^i_m) W_i}{R_{l k}}.
\end{equation}
Furthermore, the computation execution time of task $\psi^m_i$ at $CS_k$ is
\begin{equation}\small\label{eq:comTimeSj}
T^{C}_{k, i} = \frac{ (1-\mu^i_l) (1-\mu^i_n) (1-\mu^i_m) F_i}{f_{k, i}},
\end{equation}
where $f_{k, i}$ denotes $CS_k$ computation resource allocated to task $\psi^m_i$. The energy consumption of computing at $CS_k$ is expressed as
\begin{equation}\small\label{eq:energyComSj}
E^{C}_{k, i} = \iota (1-\mu^i_l) (1-\mu^i_n) (1-\mu^i_m) {F_i} f^2_{k, i}.
\end{equation}
In addition, considering that the size of computation results is significantly smaller than that of task data, and the downlink rate is much higher than the uplink rate in SAGIN, the delay associated with the transmission of computation results can be neglected in this work \cite{9344666,9374102}. Consequently, the latency of task $\psi^m_i$ is
\begin{equation}\small\label{eq:overallLantency}
\begin{aligned}
T^m_i =&~{\rm max} (T^{C}_{m, i}, T^{T}_{m, n} + T^{C}_{n, i}, T^{T}_{m, n} + T^{T}_{n, l} + T^{C}_{l, i}, T^{T}_{m, n} + T^{T}_{n, l}\\
       &+ T^{T}_{j, i} + T^{C}_{j, i}, T^{T}_{m, n} + T^{T}_{n, l} + T^{T}_{k, i} + T^{C}_{k, i}).
\end{aligned}
\end{equation}
Moreover, the total energy cost of task $\psi^m_i$ can be expressed as
\begin{equation}\small\label{eq:overallEnergyCost}
\begin{aligned}
E^m_i = &~E^{C}_{m, i} + E^{T}_{m, n, i} + E^{C}_{n, i} + E^{T}_{n, l, i} + E^{C}_{l, i}+ E^{T}_{l, j, i} + E^{C}_{j, i}\\
        &+ E^{T}_{l, k, i} + E^{C}_{k, i}.
\end{aligned}
\end{equation}

\subsection{Problem Formulation}
In this paper, our objectives are twofold: 1) To minimize the overall energy consumption in the proposed SAGIN while satisfying the latency constraint. 2) To minimize the average delay for all tasks in the proposed SAGIN with the energy consumption constraint. Thus, we have two cases:

~\emph{Case 1} - We aim to minimize the overall energy consumption in the proposed SAGIN while satisfying the latency constraint, the index of tasks from all ground users are denoted as $i \in {\mathcal I} = \{1, 2, ..., I\}$, the optimization problem can be formulated as
\begin{align}\small
    \bold{\rm (P1)}: &\min_{\mathbb{Q}, \mathbb{Z}, \boldsymbol{\rm \mu}, \mathbb{F}_U, \mathbb{F}_S, \mathbb{F}_{CS}} \sum_{m=1}^{M}\sum_{i=1}^{I} E^m_i,\label{SecrecyFunc}\\
    {\rm s.t.}&~ T_n \leq T_{U_n S_l}, \forall n \in {\mathcal N} \tag{\ref{SecrecyFunc}{a}}, \label{SecrecyFuncSuba}\\
    &\sum_{n=1}^{N} z_{m, n} = 1, \forall m \in {\mathcal M} \tag{\ref{SecrecyFunc}{b}}, \label{SecrecyFuncSubb}\\
    &\sum_{l=1}^{L} z_{n, l} = 1, \forall n \in {\mathcal N} \tag{\ref{SecrecyFunc}{c}}, \label{SecrecyFuncSubc}\\
    &\sum_{j=1,j\neq l}^{L} \sum_{k=1}^{K} z_{l, j} + z_{l, k}  = 1, \forall l \in {\mathcal L}   \tag{\ref{SecrecyFunc}{d}}, \label{SecrecyFuncSubd}\\
    &\sum_{i=1}^{I} f_{n, i} \leq f_{{\rm max}, U}, \forall n \in {\mathcal N}  \tag{\ref{SecrecyFunc}{e}}, \label{SecrecyFuncSube}\\
    &\sum_{i=1}^{I} f_{l, i} \leq f_{{\rm max}, S}, \forall l \in {\mathcal L} \tag{\ref{SecrecyFunc}{f}}, \label{SecrecyFuncSubf}\\
    &\sum_{i=1}^{I} f_{k, i} \leq f_{{\rm max}, CS}, \forall k \in {\mathcal K} \tag{\ref{SecrecyFunc}{g}}, \label{SecrecyFuncSubg}\\
    &\sum_{i=1}^{I} T^m_i \leq T_{\rm max}, \forall m \in {\mathcal M} \tag{\ref{SecrecyFunc}{h}}, \label{SecrecyFuncSubh}\\
    &x_{\rm min} \leq x_{n}(t) \leq x_{\rm max}, y_{\rm min} \leq y_{n}(t) \leq y_{\rm max},\notag\\&z_{\rm min} \leq z_{n}(t) \leq z_{\rm max}, \forall n \in {\mathcal N}~and~ \forall t \in {\mathcal T} \tag{\ref{SecrecyFunc}{i}}, \label{SecrecyFuncSubi}
\end{align}
where $\mathbb{Q} = \{Q_1, Q_2, ..., Q_N\}$, $Q_n = \{q_n(t)\}_{n \in {\mathcal N}}$ denotes the UAV $U_n$'s trajectory for dynamic access capability in MEC,  $\mathbb{Z} = \{z_{m, n}, z_{n, l}, z_{l, j}, z_{l, k}\}_{m \in {\mathcal M}, n \in {\mathcal N}, l \in {\mathcal L}, j \in {\mathcal L}, k \in {\mathcal K}}$ denotes the user/UAV/LEO pairing, $\boldsymbol{\rm \mu} = \{\mu^i_n, \mu^i_m, \mu^i_l\}_{i \in {\mathcal I}, m \in {\mathcal M}, n \in {\mathcal N}, l \in {\mathcal L}}$ denotes the partial offloading decision, $\mathbb{F}_U = \{f_{n, i}\}_{n \in {\mathcal N}, i \in {\mathcal I}}$ denotes the computation resource allocation in UAVs, $\mathbb{F}_S = \{f_{l, i}\}_{l \in {\mathcal L}, i \in {\mathcal I}}$ denotes the computation resource allocation in LEO satellites, $\mathbb{F}_{CS}= \{f_{k, i}\}_{k \in {\mathcal K}, i \in {\mathcal I}}$ denotes the computation resource allocation in cloud servers, $T_n$ represents the flight time of the UAV $U_n$ while participating in MEC, $I$ denotes the number of tasks from each ground user, $T_{\rm max}$ denotes the latency threshold for tasks from each ground user in the MEC. \eqref{SecrecyFuncSuba} presents that the flight duration of UAV $U_n$ needs to be within the corresponding LEO coverage time, \eqref{SecrecyFuncSubb} indicates that each user can only access to one UAV for offloading, \eqref{SecrecyFuncSubc} shows that each UAV can only access to one LEO for offloading, \eqref{SecrecyFuncSubd} denotes each LEO satellite only offloads the rest of a given task once. \eqref{SecrecyFuncSube}, \eqref{SecrecyFuncSubf} and \eqref{SecrecyFuncSubg} presents the computation resource allocation for UAVs, LEO satellites and cloud servers, respectively. \eqref{SecrecyFuncSubh} presents the latency constraint. \eqref{SecrecyFuncSubi} is the flying constraints for UAVs.

~\emph{Case 2} - We aim to minimizes the average latency for MEC in the proposed SAGIN while satisfying the energy consumption constraint, the optimization problem can be formulated as
\begin{align}
    \bold{\rm (P2)}: & \min_{\mathbb{Q}, \mathbb{Z}, \boldsymbol{\rm \mu}, \mathbb{F}_U, \mathbb{F}_S, \mathbb{F}_{CS}} \frac{1}{I}\sum_{m=1}^{M}\sum_{i=1}^{I} T^m_i ,\label{SecrecyFunc2}\\
    {\rm s.t.}&~ \sum_{i=1}^{I} E^m_i \leq E_{\rm max} \tag{\ref{SecrecyFunc2}{a}}, \label{SecrecyFunc2Suba}\\
    &\eqref{SecrecyFuncSuba}-\eqref{SecrecyFuncSubg}, \eqref{SecrecyFuncSubi} \notag
\end{align}
where $E_{\rm max}$ denotes the energy consumption threshold. Considering the coupled variables, non-linearity, and discrete constraints in \eqref{SecrecyFunc} and \eqref{SecrecyFunc2}, these two problems are extremely complicated non-convex optimization problems that are intractable by traditional optimization algorithms. Moreover, the high computational complexity of traditional optimization is a significant issue for the real-time computational demands in dynamic scenarios. To address those problems, and considering that \eqref{SecrecyFunc} and \eqref{SecrecyFunc2} involve long-term optimization, we introduce a multi-agent DRL algorithm in the following section. DRL is particularly suitable for tackling long-term optimization and non-convex problems, it is a pre-trained machine learning method which can adapt to different dynamic environment variables for the proposed network and requires very small computational cost during real-time prediction. Moreover, to tackle the challenge posed by the hybrid discrete and continuous action space in DRL from the proposed SAGIN, we employ an action decoupling method which involves training multiple agents and making optimization decisions based on the collective outcomes of all agents. By decoupling the actions, we enable each agent to focus on a specific aspect of the optimization problem, allowing more effective training and decision-making processes. In addition, the decision-assistant is employed to mitigate the influence of the unavailable action space by using the priori information to enhance the convergence performance.

\section{Decision-Assisted Hybrid Action Space DRL-Based Optimization} \label{sec:SAC}
Considering the long-term optimization objectives associated with these two problems, DRL is particularly well-suited to design solutions for them. However, these problems encompass intricate hybrid discrete-continuous action space challenges that pose a significant hurdle for DRL. Typically, the approaches to this issue involves either the discretization of continuous actions or the transformation of discrete actions into continuous ones. However, these strategies often lead to problems of huge action spaces or large exploration ranges. To counteract this, we propose a hybrid action space framework that effectively decouples hybrid actions and delegates them to different agents for cooperative learning. Subsequently, we implement an efficient and robust hybrid action method to amalgamate discrete and continuous policies from different agents, thereby addressing the hybrid action space problems in the proposed issues. In addition, considering the detrimental effect of unavailable actions during the training process for deep neural networks, we introduce the decision assistant method which assists the agents in differentiating the effects of various actions at the training process level, and enhancing convergence performance.

\subsection{SAC Algorithm for MEC in SAGIN}\label{sec:SAC_SAC}
Before using DRL algorithms, we need to form each proposed problem as a Markov decision process (MDP). A MDP consists of three key elements, named state, action and reward. State is utilized to describe the features of the environment. The state $s(t)$ at time slot $t$ can be expressed as
\begin{equation}\small\label{eq:state}
\begin{aligned}
s(t) =&~\{\{h_{m, n}(t)\}_{m \in \mathcal M, n \in \mathcal N}, \{T^R_{U S_l}(t)\}_{l \in \mathcal L},\\
     &~\{T^R_{CS_k S_l}(t)\}_{k \in \mathcal K, l \in \mathcal L}, \{F^{\varpi}_i(t)\}_{\varpi \ in \{M, N, L, K\}, i \in \mathcal I}\},
\end{aligned}
\end{equation}
where $\{h_{m, n}(t)\}_{m \in \mathcal M, n \in \mathcal N}$ describes the dynamic communication environment for ground user-UAVs, $\{T^R_{U S_l}(t)\}_{l \in \mathcal L}$ and $\{T^R_{CS_k S_l}(t)\}_{k \in \mathcal K, l \in \mathcal L}$ represents the remain service time of LEO satellites for UAVs and cloud servers, respectively. $\{F^{\varpi\}_i(t)}_{\varpi \in \{M, N, L, K\}, i \in \mathcal I}\}$ denotes the remain part of the computational task $\psi^m_i$ at MEC computational unit $\varpi$, where $\varpi \in \{M, N, L, K\}$. Moreover, we define the action $a(t)$ at time slot $t$ as
\begin{equation}\small\label{eq:action}
\begin{aligned}
a(t) = \{\mathbb{Q}(t), \mathbb{Z}(t), \boldsymbol{\rm \mu}(t), \mathbb{F}_U(t), \mathbb{F}_S(t), \mathbb{F}_{CS}(t)\},
\end{aligned}
\end{equation}
which considers all the optimization variables in \eqref{SecrecyFunc} and \eqref{SecrecyFunc2}. Furthermore, we also need to design the reward in MDP for the DRL algorithm. For the two cases considered in this work, we need to design two types of reward for each case. For Case 1 in \eqref{SecrecyFunc}, the reward function can be designed as
\begin{equation}\small\label{eq:reward1}
\begin{aligned}
r\big(s(t), a(t)\big) = \frac{1}{\sum_{i=1}^{I}E_i(t)}.
\end{aligned}
\end{equation}
As the objective of DRL is to maximize cumulative rewards, which contradicts our goal of minimizing overall energy consumption, we adopt the reciprocal of overall energy consumption as the reward to guide the agents in DRL towards convergence while focusing on energy minimization. On the other hand, for Case 2 in \eqref{SecrecyFunc2}, the reward function can be designed as
\begin{equation}\small\label{eq:reward2}
\begin{aligned}
r\big(s(t), a(t)\big) = \frac{1}{\sum_{i=1}^{I} T_i}.
\end{aligned}
\end{equation}
Similar to Case 1, considering the uneven distribution of energy values, negative functions may easily lead to a high variance in reward distribution, we utilize the reciprocal of the target, which is the reciprocal of latency, as the reward to encourage the agent to minimize overall latency. To optimize the variables in accordance with the aforementioned MDP for \eqref{SecrecyFunc} and \eqref{SecrecyFunc2}, we employ the SAC algorithm to explore potential solutions. The SAC algorithm is an actor-critic algorithm that combines both policy-based and value-based approaches, it consists of four neural networks: a network ${\varphi}$ representing the value function, a soft Q-network ${\theta}$ representing the action-value function, a policy network $\vartheta$ representing the policy, and a target network $\bar{\varphi}$ for the value function. Therefore, in practice we need to train three networks to form the SAC algorithm. The update function for the value function is expressed as
\begin{equation}\small\label{eq:valueFunction}
\begin{aligned}
J_{Q_{\varphi}} =&~{\mathbb E}_{s(t) \sim D} \bigg[ \frac{1}{2} \Big( Q_{\varphi}\big(s(t)\big) - {\mathbb E}_{a(t) \sim Q_\vartheta}\big[Q_{\theta}\big(s(t), a(t)\big)\\
                 &- {\rm log}Q_\vartheta \big(a(t)|s(t)\big)\big]\Big)^2 \bigg],
\end{aligned}
\end{equation}
where $Q_{\cdot}$ presents the $Q$-value from network $\cdot$, ${\mathbb E}[\cdot]$ denotes the expectation, $D$ is the distribution of the samples. In SAC, the loss is minimized by using the squared residual error in the value function. We can obtain the gradient of \eqref{eq:valueFunction} as
\begin{equation}\small\label{eq:valueGradient}
\begin{aligned}
\bar{\nabla}_{\varphi} J_{Q_{\varphi}} =&~\nabla_{\varphi} Q_{\varphi}\big(s(t)\big) \bigg( Q_{\varphi}\big(s(t)\big) - Q_{\theta}\big(s(t), a(t)\big)\\
                                        &+ {\rm log}Q_\vartheta \big(a(t)|s(t)\big)\bigg).
\end{aligned}
\end{equation}

Moreover, the update function of the soft Q-network $Q_{\theta}$ can be expressed as
\begin{equation}\small\label{eq:SoftQFunction}
\begin{aligned}
J_{Q_{\theta}} = {\mathbb E}_{\big(s(t), a(t)\big) \sim D} \bigg[ \frac{1}{2} \Big(Q_{\theta}\big(s(t), a(t)\big) - \bar{Q}\big(s(t), a(t)\big) \Big)^2  \bigg],
\end{aligned}
\end{equation}
where $\bar{Q}\big(s(t), a(t)\big)$ is given by
\begin{equation}\small\label{eq:SoftQFunctionQvalue}
\begin{aligned}
\bar{Q}\big(s(t), a(t)\big) = r\big(s(t), a(t)\big) + \tau{\mathbb E}_{s(t+1) \sim \varrho} \Big[Q_{\bar{\varphi}}\big(s(t+1)\big)\Big],
\end{aligned}
\end{equation}
where $\tau$ is the discount factor in the SAC, $\varrho$ denotes the transition probability. The gradients could be calculated as
\begin{equation}\small\label{eq:SoftQGradient}
\begin{aligned}
\bar{\nabla}_{\varphi} J_{Q_{\theta}} = &~{\nabla}_{\varphi}Q_{\theta}\big(s(t), a(t)\big) \Big(Q_{\theta}\big(s(t), a(t)\big) - r\big(s(t), a(t)\big)\\
                                        &- \tau Q_{\bar{\varphi}}\big(s(t+1)\big)\Big).
\end{aligned}
\end{equation}

In addition, we reconfigure the policy network via a neural network transformation
\begin{equation}\small\label{eq:PolicyReparameteri}
\begin{aligned}
a(t) = f_{\vartheta}(\varsigma(t); s(t)),
\end{aligned}
\end{equation}
where $\varsigma$ is a noise term that follows Gaussian distribution. Thus, the update function of the policy network $Q_{\vartheta}$ can be expressed as
\begin{equation}\small\label{eq:PolicyFunctionNoise}
\begin{aligned}
J_{Q_{\vartheta}} = &~{\mathbb E}_{s(t) \sim D, \varsigma(t) \sim \phi}\bigg[{\rm log}Q_{\vartheta}(f_{\vartheta}\big(\varsigma(t); s(t))|s(t)\big)\\
                    &- Q_{\theta}\big(s(t), f_{\vartheta}(\varsigma(t); s(t))\big) \bigg],
\end{aligned}
\end{equation}
where $\phi$ denotes the normal distribution. The gradient can be calculated as
\begin{equation}\small\label{eq:PolicyFunctionGradient}
\begin{aligned}
\bar{\nabla}_{\varphi} J_{Q_{\vartheta}} = &~{\nabla}_{\vartheta}{\rm log}Q_{\vartheta}\big(a(t)|s(t)\big) + \bigg({\nabla}_{a(t)}{\rm log}Q_{\vartheta}\big(a(t)|s(t)\big)\\
                                           &- {\nabla}_{a(t)}Q\big(s(t), a(t)\big) \bigg) {\nabla}_{\vartheta}f_{\vartheta}(\varsigma(t); s(t)),
\end{aligned}
\end{equation}
where
\begin{equation}\small\label{eq:PolicyFunctionGradientQvalue}
\begin{aligned}
Q\big(s(t), a(t)\big) \leftarrow ~ &~r_{\vartheta}\big(s(t), a(t)\big)\\
                                   &+ \tau{\mathbb E}_{s(t+1) \sim \varrho, a(t+1) \sim \vartheta} \Big[ \bar{Q}_{\varphi}\big(s(t+1)\big)\Big].
\end{aligned}
\end{equation}

\subsection{Hybrid Action Space SAC Algorithm}\label{sec:SAC_Hybrid}
In the proposed problems \eqref{SecrecyFunc} and \eqref{SecrecyFunc2}, the decisions are hybrid discrete and continuous which poses a challenge to existing DRL algorithms \cite{neunert2020continuous,XU2023141}. Many existing methods directly approximate the continuous action space to discrete space, resulting in a significantly larger discrete action space. Another approach is to transform discrete actions into continuous ones, which introduces greater complexity and difficulty in achieving convergence during training. Therefore, we introduce an action decoupling algorithm in SAC to address this challenge. The action and policy are redefined as
\begin{align}
   A &~= \{\emptyset, \partial \}, \label{AD_newAction}\\
   Q_\vartheta(a|s) &~= Q^{\emptyset}_\vartheta(a^\emptyset|s)Q^{\partial}_\vartheta(a^\partial|s)\\
   &= \prod_{a_i \ in a^\emptyset}Q^{\emptyset}_\vartheta(a_i|s) \prod_{a_j \ in a^\partial}Q^{\partial}_\vartheta(a_j|s), \label{AD_newPolicy}
\end{align}
where $\emptyset$ denotes the discrete actions, $\partial$ denotes the continuous actions. Therefore, we design two agents, each equipped with an independent SAC framework, responsible for adjusting the discrete or continuous optimization variables from \eqref{SecrecyFunc} and \eqref{SecrecyFunc2}. Thus, we have the cost functions of value networks for discrete and continuous agents as
\begin{equation}\small\label{AD_newLoss1}
\begin{aligned}
    J_{Q^{\emptyset}_{\varphi}} =&~{\mathbb E}_{s(t) \sim D} \bigg[ \frac{1}{2} \Big( Q^{\emptyset}_{\varphi}\big(s(t)\big) - {\mathbb E}_{a(t) \sim Q^{\emptyset}_\vartheta}\big[Q^{\emptyset}_{\theta}\big(s(t), a(t)\big)\\
    &- {\rm log}Q^{\emptyset}_\vartheta \big(a(t)|s(t)\big)\big]\Big)^2 \bigg],
\end{aligned}
\end{equation}
\begin{equation}\small\label{AD_newLoss2}
\begin{aligned}
    J_{Q^{\partial}_{\varphi}} =&~{\mathbb E}_{s(t) \sim D} \bigg[ \frac{1}{2} \Big( Q^{\partial}_{\varphi}\big(s(t)\big) - {\mathbb E}_{a(t) \sim Q^{\partial}_\vartheta}\big[Q^{\partial}_{\theta}\big(s(t), a(t)\big)\\
    &- {\rm log}Q^{\partial}_\vartheta \big(a(t)|s(t)\big)\big]\Big)^2 \bigg].
\end{aligned}
\end{equation}
Furthermore, we also have the cost functions of soft-Q-networks for discrete and continuous agents as
\begin{equation}\small\label{AD_newLoss3}
\begin{aligned}
    J_{Q^{\emptyset}_{\theta}} = {\mathbb E}_{\big(s(t), a(t)\big) \sim D} \bigg[ \frac{1}{2} \Big(Q^{\emptyset}_{\theta}\big(s(t), a(t)\big) - \bar{Q}^{\emptyset}\big(s(t), a(t)\big) \Big)^2  \bigg],
\end{aligned}
\end{equation}
\begin{equation}\small\label{AD_newLoss4}
\begin{aligned}
    J_{Q^{\partial}_{\theta}} = {\mathbb E}_{\big(s(t), a(t)\big) \sim D} \bigg[ \frac{1}{2} \Big(Q^{\partial}_{\theta}\big(s(t), a(t)\big) - \bar{Q}^{\partial}\big(s(t), a(t)\big) \Big)^2  \bigg].
\end{aligned}
\end{equation}
Moreover, we can obtain the cost functions of policy networks for discrete and continuous agents as
\begin{equation}\small\label{AD_newLoss5}
\begin{aligned}
    J_{Q^{\emptyset}_{\vartheta}} =&~{\mathbb E}_{s(t) \sim D, \varsigma(t) \sim \phi}\bigg[{\rm log}Q^{\emptyset}_{\vartheta}(f_{\vartheta}\big(\varsigma(t); s(t))|s(t)\big)\\
    &- Q^{\emptyset}_{\theta}\big(s(t), f_{\vartheta}(\varsigma(t); s(t))\big) \bigg],
\end{aligned}
\end{equation}
\begin{equation}\small\label{AD_newLoss6}
\begin{aligned}
    J_{Q^{\partial}_{\vartheta}} =&~{\mathbb E}_{s(t) \sim D, \varsigma(t) \sim \phi}\bigg[{\rm log}Q^{\partial}_{\vartheta}(f_{\vartheta}\big(\varsigma(t); s(t))|s(t)\big)\\
    &- Q^{\partial}_{\theta}\big(s(t), f_{\vartheta}(\varsigma(t); s(t))\big) \bigg].
\end{aligned}
\end{equation}

\begin{figure}[t!]
  \centering
  \centerline{\includegraphics[width=0.42\textwidth]{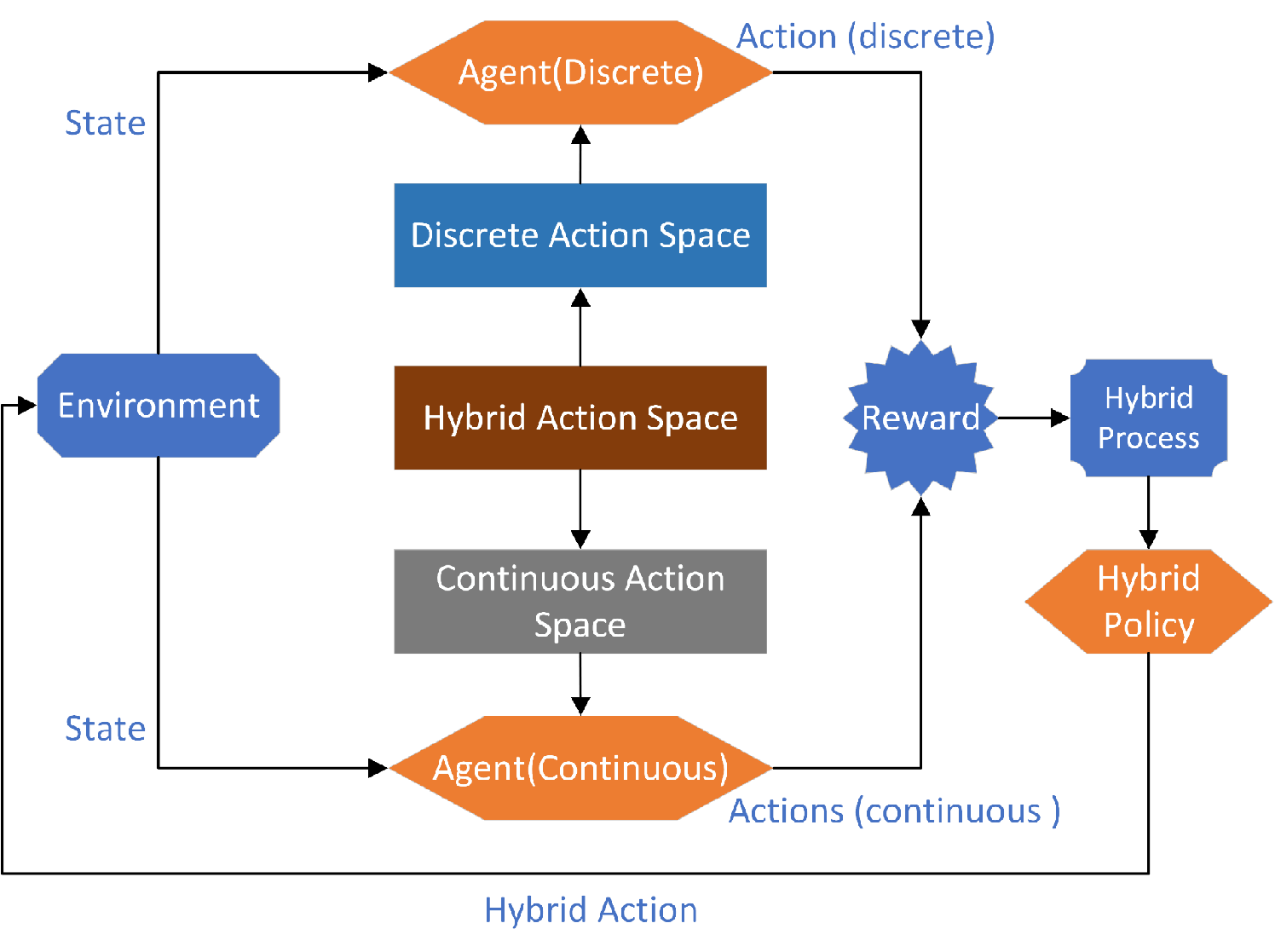}}
 \caption{The framework of the hybrid action space SAC algorithm.} \label{fig:hybrid}
\end{figure}

Subsequently, we employed the maximum aposteriori policy optimisation \cite{abdolmaleki2018maximum} algorithm to jointly optimize the decoupled policies. We assume a new policy $\ell$ for the hybrid action space SAC, the update function of $\ell$ is given as
\begin{equation}\small\label{eq:ellUpdate}
\begin{aligned}
J_{Q_{\ell}} ={\mathbb E}_{\ell(a|s)}[\hat{Q}(s, a)],
\end{aligned}
\end{equation}
where $\hat{Q}$ is the Q-function learned from the replay $D$ \cite{neunert2020continuous}. It is crucial to ensure that the new policy does not deviate significantly from the current one. Thus, we have
\begin{equation}\small\label{eq:ellUpdateKL}
\begin{aligned}
{\mathbb E}_{s \sim D}[\rho(Q_{\ell}(a|s)||Q_{\bar{\ell}}(a|s))] < \chi,
\end{aligned}
\end{equation}
where $\rho(\cdot)$ denotes the Kullback-Leibler divergence, $\bar{\ell}$ denotes the current hybrid policy, $\chi$ denotes the stability threshold. Then we update the new policy as
\begin{align}
\hat{\ell} = ~&\arg \max_{\ell}{\mathbb E}_{s \sim D}[\rho(Q_{\ell}(a|s)||Q^{\emptyset}_{\ell}(a^\emptyset|s)Q^{\partial}_{\ell}(a^\partial|s))], \label{ellUpdate2}\\
{\rm s.t.}&~{\mathbb E}_{s \sim D}[\rho(Q^{\emptyset}_{\bar{\ell}}(a^\emptyset|s)||Q^{\emptyset}_{\ell}(a^\emptyset|s))] < \chi_{\emptyset},\tag{\ref{ellUpdate2}{a}}, \label{ellUpdate2a}\\
 &~{\mathbb E}_{s \sim D}[\frac{1}{\bar{o}}\sum_{o=1}^{\bar{o}} \rho(Q^{\partial}_{\bar{\ell}}(a^\partial|s)||Q^{\partial}_{\ell}(a^\partial|s))] < \chi_{\partial} \tag{\ref{ellUpdate2}{b}}, \label{ellUpdate2b}
\end{align}
where $\chi_{\emptyset}$ and $\chi_{\partial}$ are the stability thresholds for discrete and continuous agents, respectively, $\bar{o}$ denotes the number of discrete choices. The framework of the hybrid action space SAC algorithm is shown in Fig. \ref{fig:hybrid}.

\emph{Remark 1}: The proposed algorithm decouples the hybrid action space into discrete and continuous actions, assigning them to different agents, each trained with its independent SAC framework. Subsequently, the maximum aposteriori policy optimization algorithm is employed to merge the decoupled policies from all agents. The proposed algorithm addresses the challenges of the hybrid action space in the proposed problems and provides a flexible framework for hybrid action selection.

\subsection{Decision-Assisted DRL}\label{sec:SAC_DA}
In Section \ref{sec:SAC_SAC}, when designing the reward function for the SAC algorithm, each action is indiscriminately given a corresponding reward, without considering whether the action itself is available. For example, a certain MEC unit, such as the UAV $U_n$, has already completed its allocated portion of computing task $\psi^m_i$. In this case, the action of continuing to allocate CPU cycles to this task should be deemed as unavailable. Similarly, if a cloud server falls outside the coverage range of a LEO satellite $S_l$ and consequently can no longer connect to it, the action of $S_l$ still choosing this cloud server as an offloading target should also be deemed unavailable. In these situations, DRL needs a long time to learn the mutual impact between different state-action pairs, leading to a low convergence efficiency (even inability to converge). To solve this problem, a common solution in DRL-related studies is to introduce negative rewards, using them to penalize ineffective actions or those that deviate from the target, thereby speeding up convergence.

\begin{figure}[t!]
  \centering
  \centerline{\includegraphics[width=0.37\textwidth]{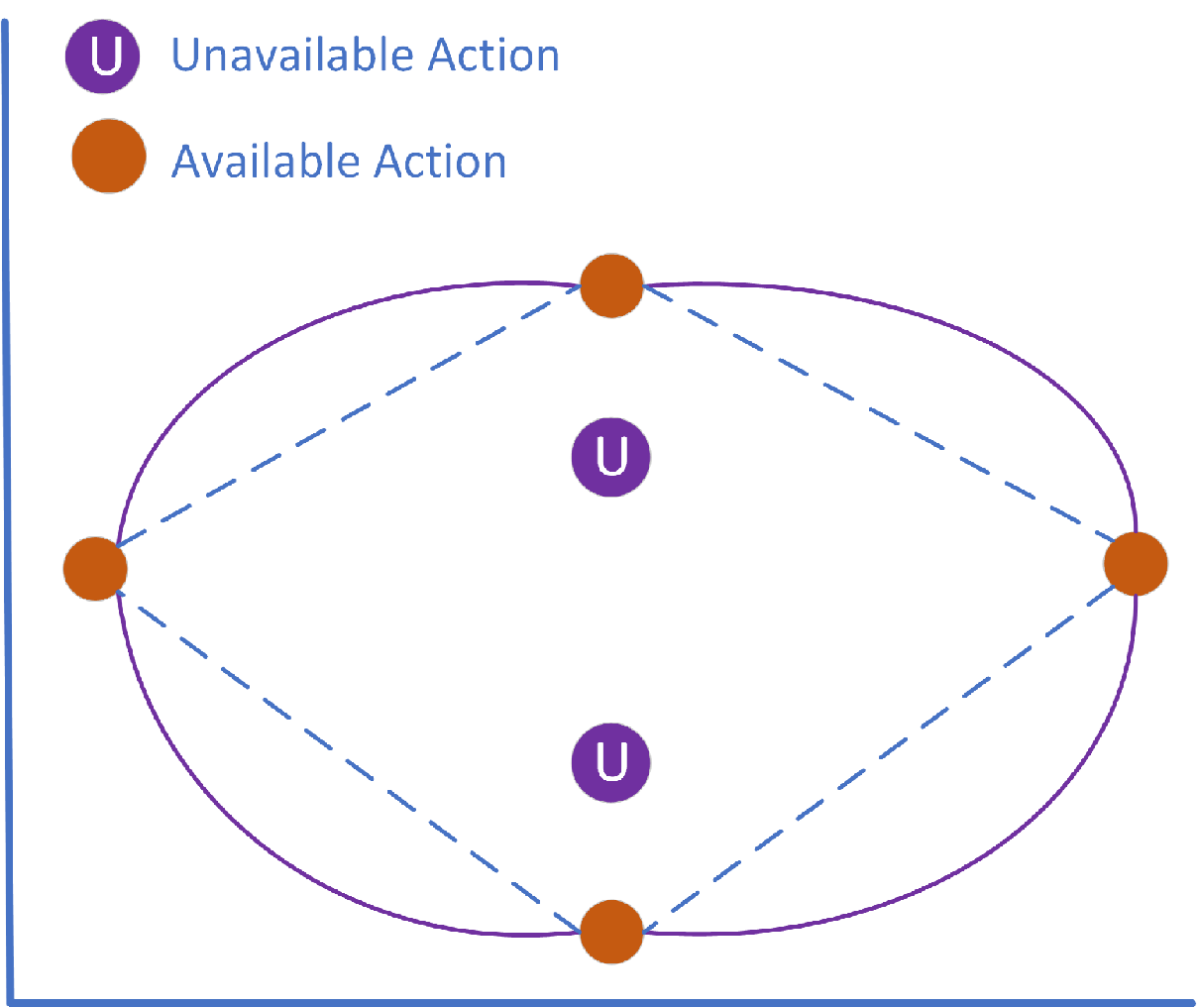}}
 \caption{The impact of unavailable actions on the training of deep neural networks for function approximation.} \label{fig:UA}
\end{figure}

However, when employing a negative reward mechanism, the punitive aspect might be overly harsh, making it difficult for the agent to properly differentiate actions that receive no reward or very small positive reward from those that receive other positive rewards. Under randomly initialized weights, the neural network might not provide sufficient rewards for the optimal action after several iterations. As a result, the negative reward could lead the agent to converge more easily to local optima \cite{9321455}. An alternative solution is to directly prohibit the agent from choosing unavailable actions. However, whether the actions are available or not, they all originate from the same policy. This policy can only indicate the advantages and disadvantages of state-action pairs, but it cannot differentiate available and unavailable actions. Therefore, even ineffective actions can influence the neural network that generates the policy. As shown in Fig. \ref{fig:UA}, when disregarding unavailable actions, the outcomes can be represented by the dashed lines in the graph, akin to drawing direct connections between available actions. However, when accounting for unavailable actions, the penalization mechanism for such actions pushes the fitted line as far away from them as possible, causing it to deviate from its original position and resulting in the effect depicted by the solid lines in the graph. Therefore, ignoring unavailable actions directly can also have a negative impact on the learning outcome in the DRL. In the SAGIN system proposed, if the LEO satellites that have already exceeded their coverage range are simply disregarded, the function approximator capability of the neural network may still lead to an influence on the resource allocation choices of other LEO satellites.

\begin{figure}[t!]
  \centering
  \centerline{\includegraphics[width=0.44\textwidth]{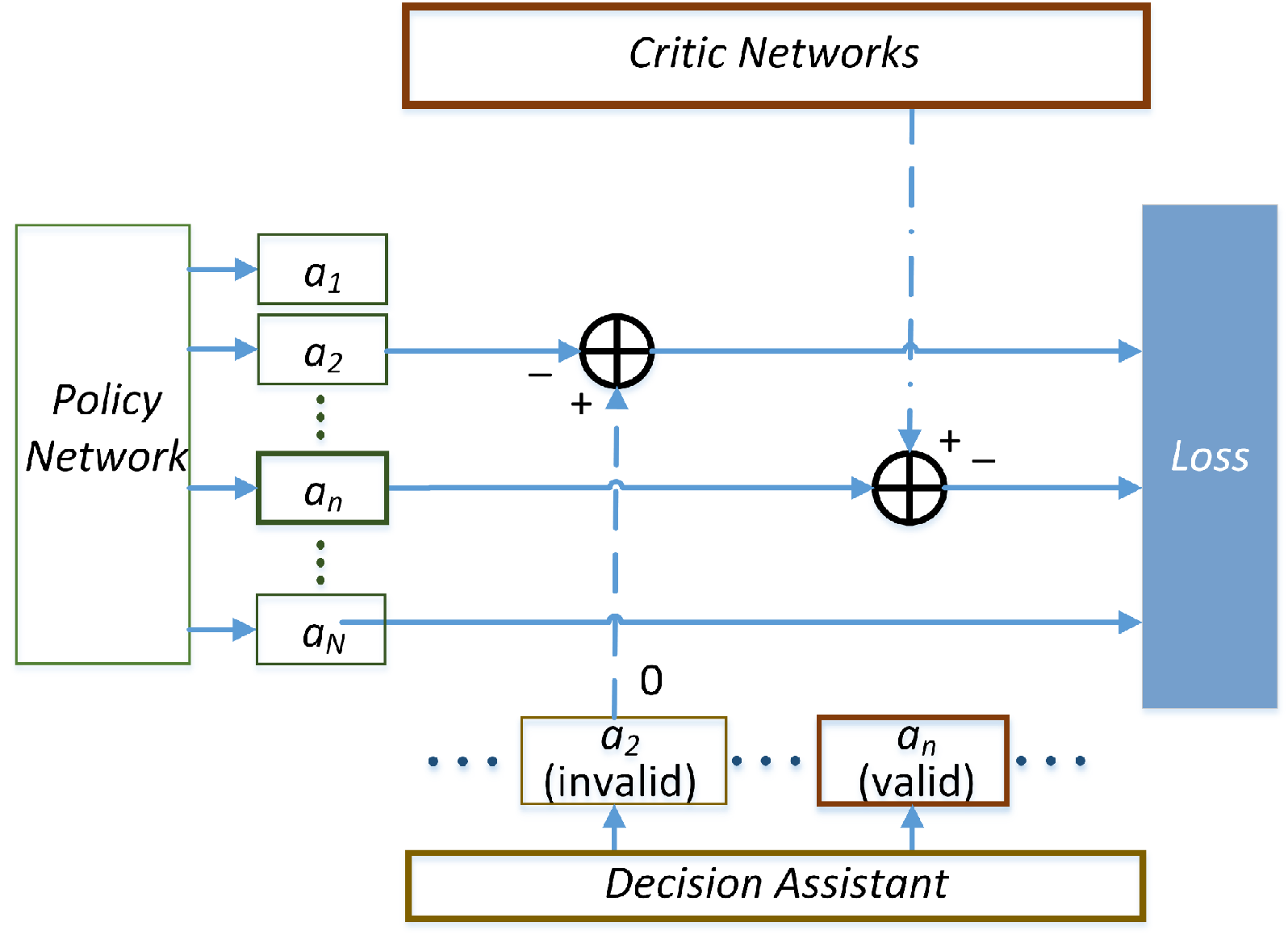}}
 \caption{The framework of the decision assisted DRL method.} \label{fig:DA}
\end{figure}

To address this issue, we introduce a decision-assisted DRL algorithm to mitigate the negative impact of unavailable actions on the training process of DRL. Unlike most previous works that utilize negative rewards or simply disregard unavailable actions, we directly set the weights of state-unavailable action pairs to zero. We generate decision-assisted DRL training pairs as
\begin{equation}\small\label{eq:DApair}
\begin{aligned}
\{s(t), a(t), 0\}.
\end{aligned}
\end{equation}
This is an auxiliary method that uses prior knowledge in deep learning to train deep neural networks. As shown in Fig. \ref{fig:DA}, each time the policy network is updated, not only are the weights of the valid actions updated, but invalid actions are also selected and continuously trained. During training, the neural network will continually match the minimal weight to the invalid actions, greatly diminishing their impact. In addition, the generation of invalid actions does not require an exploration process, this significantly reduces the exploration dimensions during training and improves convergence efficiency. The pseudo-code of the decision-assisted multi-agent SAC with hybrid action space (DM-SAC-H) is in Algorithm \ref{Algorithm DA_SAC_H}.

\begin{algorithm}[t!]
\caption{\textbf{DM-SAC-H}: }\label{Algorithm DA_SAC_H}
\begin{algorithmic}[1]
 \makeatletter\setcounter{ALG@line}{0}\makeatother
 \State Initialize all the networks, $\varphi^\emptyset$, $\theta^\emptyset$, $\vartheta^\emptyset$, $\bar{\varphi}^\emptyset$, $\varphi^\partial$, $\theta^\partial$, $\vartheta^\partial$, $\bar{\varphi}^\partial$ and $\ell$.
 \Repeat:
 \For {each time slot}
 \State Obtain $a(t)$, $s(t+1)$, $r\big(s(t), a(t)\big)$ based on the
 \Statex \phantom{0}\phantom{0}\phantom{0}\phantom{0}\phantom{0}\phantom{0}policy $\ell$ and current state $s(t)$.
 \State Save the sample $\{s(t), a(t), s(t+1), r\big(s(t), a(t)\big)\}$
 \Statex \phantom{0}\phantom{0}\phantom{0}\phantom{0}\phantom{0}\phantom{0}to replay $D$.
 \State Generate samples for unavailable actions and save
 \Statex \phantom{0}\phantom{0}\phantom{0}\phantom{0}\phantom{0}\phantom{0}them to replay $D$.
 \EndFor
 \For {each training epoch}
 \State Update $Q_{\varphi^\emptyset}$ based on \eqref{AD_newLoss1}, $Q_{\theta^\emptyset}$ based on \eqref{AD_newLoss3},
 \Statex \phantom{0}\phantom{0}\phantom{0}\phantom{0}\phantom{0}\phantom{0}$Q_{\vartheta^\emptyset}$ based on \eqref{AD_newLoss5} and samples from \eqref{eq:DApair}.
 \State Update $Q_{\varphi^\partial}$ based on \eqref{AD_newLoss2}, $Q_{\theta^\partial}$ based on \eqref{AD_newLoss4},
 \Statex \phantom{0}\phantom{0}\phantom{0}\phantom{0}\phantom{0}\phantom{0}$Q_{\vartheta^\partial}$ based on \eqref{AD_newLoss6} and samples from \eqref{eq:DApair}.
 \State Soft update $Q_{\bar{\varphi^\emptyset}}$ and $Q_{\bar{\varphi^\partial}}$.
 \State Update $\ell$ based on \eqref{eq:ellUpdate}, \eqref{eq:ellUpdateKL} and \eqref{ellUpdate2}.
 \EndFor
 \Until{convergence.}
\end{algorithmic}
\end{algorithm}

\section{Simulation Results} \label{sec:sim}
In the simulation, we utilized TensorFlow-2 to train the proposed algorithm. Unless otherwise stated, parameters are set as follows: the transmit power of ground users $P_G = 0.1$ W, the transmit power of UAVs $P_U = 1$ W, the transmit power of LEO satellite $P_S = 1$ W, the noise level is -100 dBm \cite{9725258}, the transmission overhead factor $b = 1.2$, the path loss exponents $\alpha_L = 2$ and $\alpha_N = 2.5$, respectively. The altitude of LEO satellite is 800 km, the frequency for UAV is 1 GHz, the beam gain $\delta_l = 25$ dB, the carrier frequency of LEO satellites is 30 GHz, the maximum Doppler frequency is referred to the Ka-band in \cite{10043628}, the normalized gain $G^d$ is 1 for all ISL \cite{9327501}, the thermal noise $\zeta = 354.81$ K \cite{9327501}, the bandwidth for ISL and satellite-Cloud is 1 GHz \cite{10043628}, other bandwidths are 10 MHz, the minimum elevation angle $\omega_G = 40 ^{\circ}$, the data size of each task is randomly generated from 0.6 MB to 1.2 MB, the required computational CPU cycles of each task is 3 Gcycles, the energy factor $\iota = 10^{-25}$, the computing capacities of each ground user, UAV, LEO satellite and cloud server are 0.1 GHz, 0.5 GHz, 1 GHz, and 3 GHz, respectively. The discount factor in DRL is 0.99, Adam optimizer is utilized in the proposed algorithm. The tasks are randomly generated by DAGs, each ground user has 10 tasks. We set the number of ground users $M = 12$, the number of UAVs $N = 3$, the number of LEO satellites $L = 5$, the number of cloud server $K = 3$, the thresholds $T_{\rm max} = 50$ s, $E_{\rm max} = 400$ J. The ground users are randomly distributed in a square of 200 m $\times$ 200 m, UAVs are randomly distributed in a square of 200 m $\times$ 200 m with the height of 60 m at the beginning, the distance of LEO satellites are randomly generated from 100 km to 500 km, the cloud servers are randomly distributed at intervals of 50 km to 100 km centered around the midpoint of the ground users. UAVs can only fly in a square of 200 m $\times$ 200 m with the minimum height of 50 m and the maximum height of 60 m. All agents interacted with the environment for approximately 100,000 episodes (each episode representing a complete multi-task computation process within the MEC framework) to achieve convergence. Moreover, we compare the proposed algorithm DM-SAC-H with the following benchmarks: Multi-agent PPO (M-PPO) \cite{MPPO} (widely used DRL algorithm), Parametrized DQN \cite{xiong2018parametrized} (a DQN \& DDPG hybrid framework), and A3C \cite{mnih2016asynchronous} (a asynchronous multi-agent DRL approach).

\begin{figure}[t!]
  \centering
  \centerline{\includegraphics[scale=0.55]{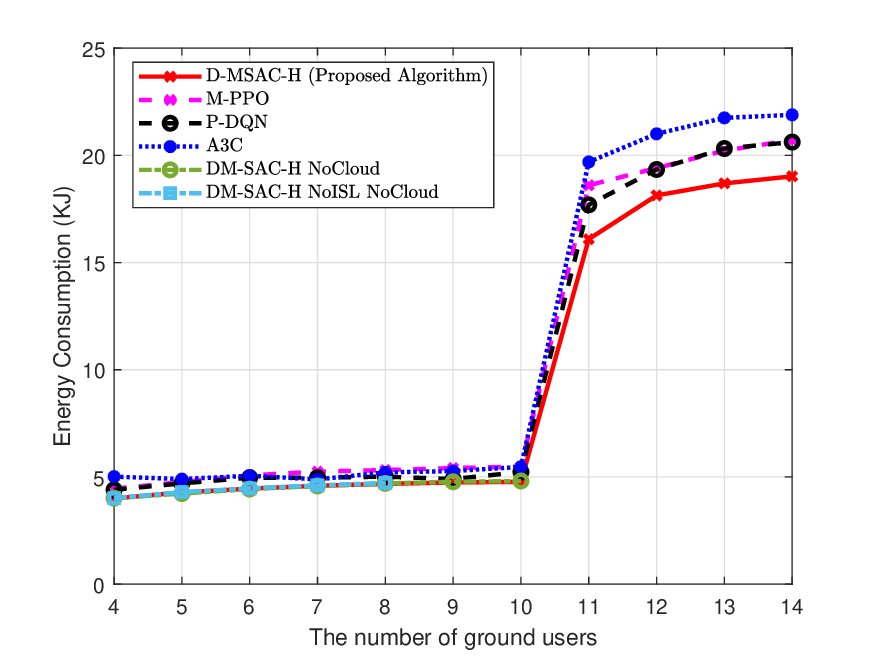}}
 \caption{\small Energy consumption versus the number of ground users in case 1.} \label{fig:R1}
\end{figure}

Fig. \ref{fig:R1} presents the energy consumption versus the number of ground users for the proposed algorithm and benchmarks in \emph{Case 1}. `DM-SAC-H NoCloud' represents the proposed method within the SAGIN framework but without the inclusion of cloud servers, `DM-SAC-H NoISL NoCloud' denotes the scenario where each LEO satellite is unable to offload tasks to other LEO satellites or cloud servers. As depicted in the results, it is clear that energy consumption rises with an increase in the number of ground users. This is primarily due to the fact that with more ground users, there is a higher demand for offloading tasks to the UAVs, LEO satellites, and cloud servers, all of which possess greater computational capacities but consume significantly more energy. Furthermore, the energy consumption escalates rapidly when $M \geq 11$; this is a consequence of the LEO satellites' inability to complete the computing task within the latency constraint, leading to the necessity of offloading the remaining tasks to the high energy-intensive cloud servers in MEC. It is also worth noting that the proposed DM-SAC-H algorithm achieves approximately 18.13 KJ when $M = 12$, in contrast to the benchmarks which obtain 19.4, 19.3 and 21 KJ, respectively. These results demonstrate that the proposed algorithm leverages the hybrid action method and decision assistant to enhance the performance of convergence. Besides, the absence of cloud servers in MEC results in the inability to meet the latency requirements when the number of ground users exceeds 10. Concurrently, in scenarios without both ISL and cloud servers, MEC fails to complete all tasks when $M \geq 8$. These findings underscore the significance of incorporating both ISL and cloud in MEC.

\begin{figure}[t!]
  \centering
  \centerline{\includegraphics[scale=0.55]{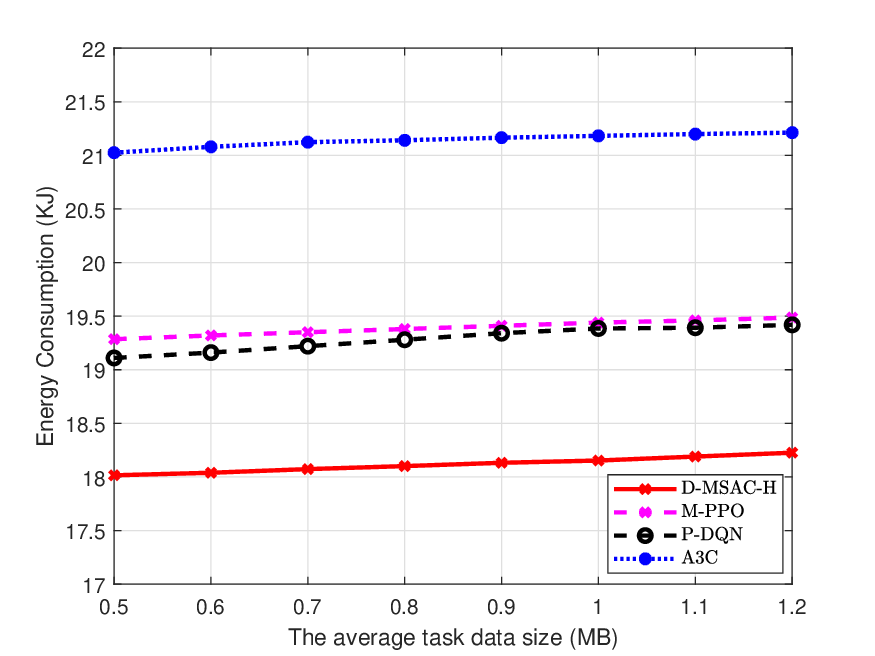}}
 \caption{\small Energy consumption versus the average task data size in case 1.} \label{fig:R2}
\end{figure}

Fig. \ref{fig:R2} indicates a comparison of energy consumption between the proposed method and the benchmarks in case 1. First, it is evident that energy consumption escalates as the average task data size increases. This is attributable to the larger task data size necessitating a more extended offloading time, as in \eqref{eq:transTimeGmUn}, \eqref{eq:transTimeUnSl}, \eqref{eq:comTimeSj} and \eqref{eq:comTimeCSk}. However, it is noteworthy that the influence of the average task data size is not substantial. This is predicated on the assumption that the required computational CPU cycles do not correlate with the data size of the task due to the diversity of tasks such as image processing and natural language processing. Although a larger task size may provide more information for computation, the computation process itself may remain unchanged. Thus, when the volume of information (task data size) increases while the required computational cycles remain constant, there is a moderate increase in energy consumption.

\begin{figure}[t!]
  \centering
  \centerline{\includegraphics[scale=0.55]{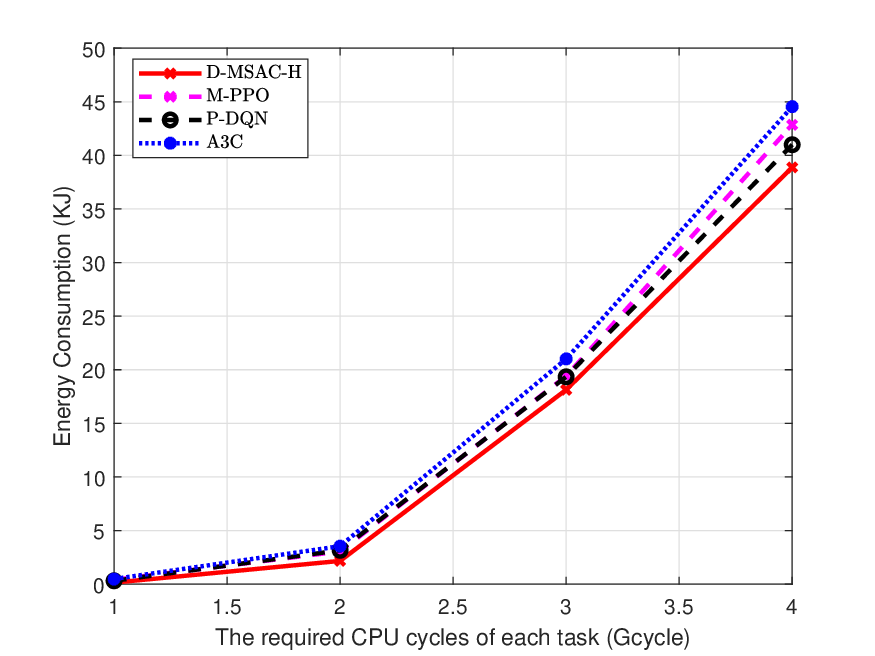}}
 \caption{\small Energy consumption versus the required CPU cycles of each task in case 1.} \label{fig:R3}
\end{figure}

As illustrated in Fig. \ref{fig:R3}, energy consumption in each instance rises as the requirement for CPU cycles per task increases. This is attributable to the necessity of offloading a larger portion of the computational tasks to the UAVs, LEO satellites, and cloud servers as the capacity of the computational task increases. Furthermore, the proposed algorithm outperforms the benchmarks; DM-SAC-H reduce the energy consumption to 2.16 KJ with 2 Gcycles, while benchmarks consume between 3 KJ to 3.5 KJ. This improved performance is due to the DM-SAC-H algorithm's combination of the hybrid action space and its usage of a priori information to mitigate the impact of unavailable actions, thereby enhancing the convergence performance. It is important to note that for computational tasks requiring over 4Gcycles, the proposed framework is unable to complete the tasks within the latency constraint due to the substantial computational requirements.

\begin{figure}[t!]
  \centering
  \centerline{\includegraphics[scale=0.55]{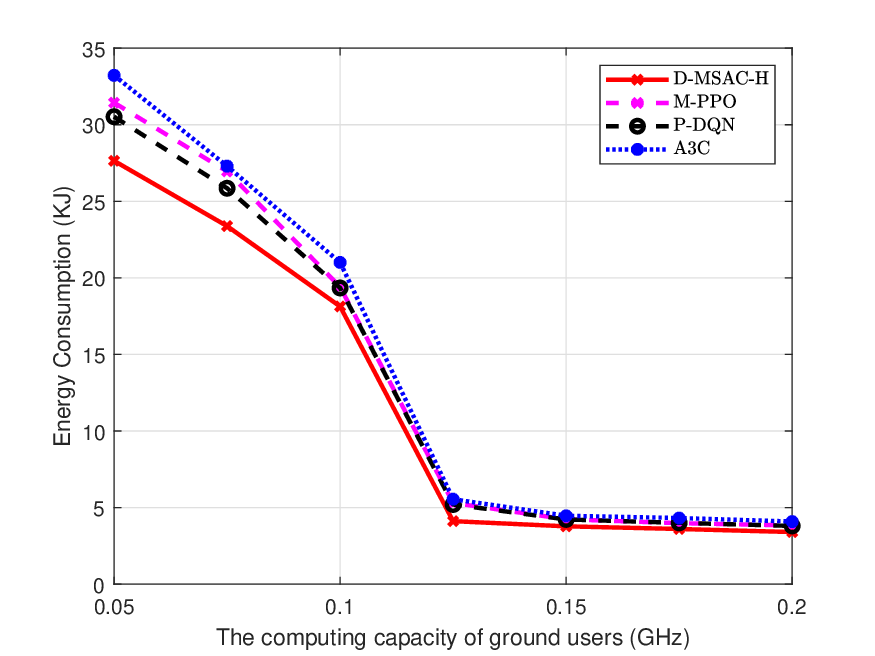}}
 \caption{\small Energy consumption versus the computing capacity of ground users in case 1.} \label{fig:R4}
\end{figure}

Fig. \ref{fig:R4} illustrates the examination of energy consumption as the computing capacity of ground users increases for both the proposed scheme and the benchmarks. It is evident that the proposed DM-SAC-H algorithm converges efficiently and surpasses all three benchmarks. When the computing capacity of ground users is 1.25 GHz, the proposed algorithm reduces energy expenditure to approximately 4.11 KJ while the other benchmarks consume over 5.2 KJ. Furthermore, when the computing capacity of ground users exceeds 1.25 GHz, there is a rapid decrease in energy consumption. This is due to the increased computing capacity enabling the ground users to complete a larger portion of the tasks, thereby only necessitating the offloading of the remainder to the UAVs and LEOs.

\begin{figure}[t!]
  \centering
  \centerline{\includegraphics[scale=0.55]{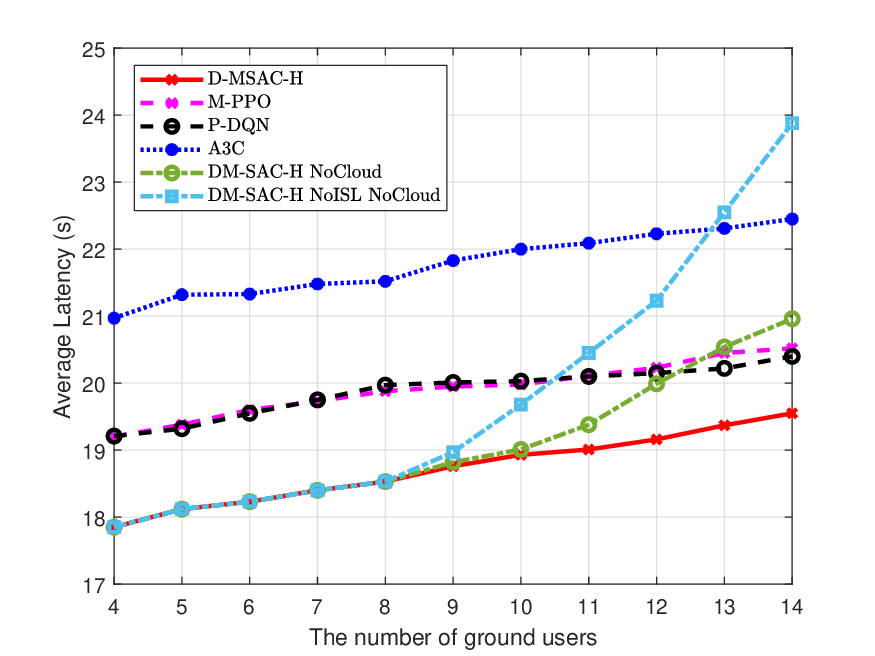}}
 \caption{\small Average latency versus the number of ground users in case 2.} \label{fig:R5}
\end{figure}

In Fig. \ref{fig:R5}, we assess the average latency of both the proposed scheme and the benchmarks in relation to different numbers of ground users. As shown in all the results, the average latency increases as the number of ground users rises. This is attributable to an increase in the number of tasks as the number of ground users expands, prompting each MEC unit to allocate computational resources to the additional tasks. Moreover, the proposed DM-SAC-H algorithm manages to reduce the average latency to 19 s with 12 ground users, whereas the other benchmarks register results exceeding 20 s. Notice that the A3C exhibits the highest latency and instability in convergence due to the low sample efficiency and the asynchronous framework. Moreover, as the number of ground users escalates, the total task count rises, resulting in a significant increase in time cost in scenarios absent of ISL or cloud servers. This outcome effectively validates the benefits of incorporating cloud servers and ISL into MEC within the SAGIN framework.

\begin{figure}[t!]
  \centering
  \centerline{\includegraphics[scale=0.55]{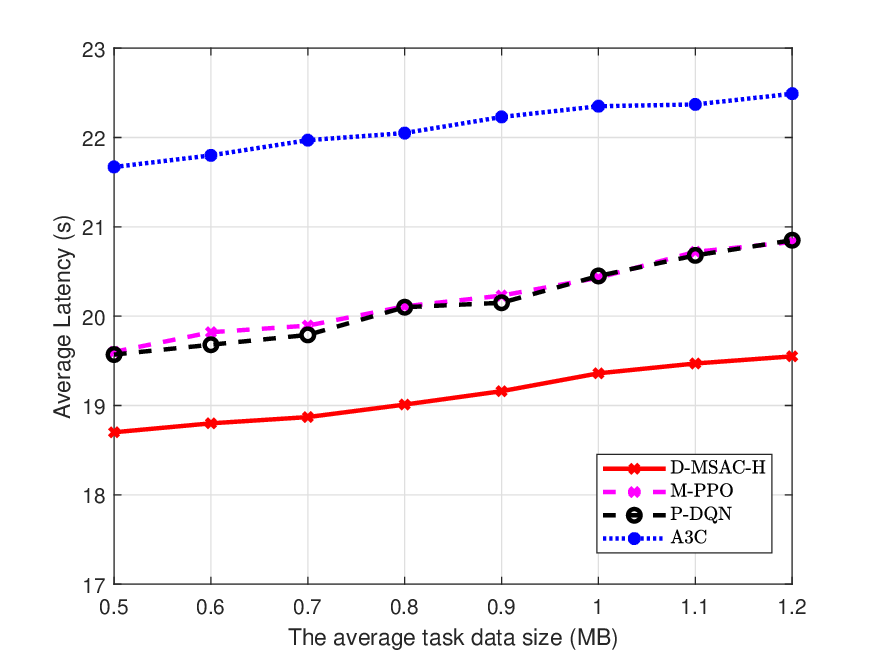}}
 \caption{\small Average latency versus the average task data size in case 2.} \label{fig:R6}
\end{figure}

Fig. \ref{fig:R6} shows a comparison of the average latency between the proposed algorithm and the benchmarks in relation to the average task data size in case 2. We can see that larger task sizes result in extended transmission times during offloading in MEC. The proposed algorithm achieves an average latency of 19.3 s when the average task size is 1 MB, while the other benchmarks yield higher latencies of 20.34 s, 20.40, and 22.31 s, respectively. These results underscore the efficacy of our proposed algorithm in utilizing the advantages of the hybrid action space and the decision assistant to achieve superior convergence.

\section{Conclusion}\label{sec:con}
In this paper, we proposed the integration of a hybrid cloud and MEC within SAGINs. This framework comprises multiple UAVs, LEO satellites and cloud servers. We considered user pairing, partial offloading, UAV trajectories and LEO satellite coverage time in this work. Moreover, we have modelled the multi-task dependency as DAGs. Aiming to minimize energy consumption under latency constraints and minimize latency under energy consumption constraints within the proposed SAGIN, we proposed a multi-agent DRL based algorithm. This learning algorithm optimizes UAV trajectories, user/UAV/LEO pairing, offloading strategy, and computational resource allocation. To tackle the challenge of the hybrid discrete and continuous action space, we introduced an action decoupling method within the multi-agent SAC algorithm to improve convergence performance. Furthermore, we employed a decision assistant to mitigate the negative impact of unavailable actions in the DRL, thereby reducing the exploration range for agents. Simulation results underscore that the proposed DM-SAC-H algorithm offers adaptable optimization solutions for the hybrid cloud and MEC within the SAGIN. When operating under latency constraints, this algorithm surpasses benchmarks by achieving a substantial reduction in total energy consumption. Conversely, under energy consumption constraints, it surpasses the benchmarks by further reducing the average latency. Furthermore, these simulation outcomes emphatically highlight the important role of cloud servers and ISLs in attaining overall optimization improvements. In our future work, we will further consider next generation multiple access (NGMA) techniques to mitigate interference between users and to enhance the resource utilization of MEC in SAGINs. In addition, we will consider more realistic scenarios such as the impact of the elevation angle in satellite-to-ground communications, and further our research on incorporating techniques such as large AI models to further strengthen the adaptive capacity of the proposed algorithm to realistic environments.

\bibliographystyle{ieeetr}
\bibliography{ref}
\end{document}